\documentclass[aps,pre,onecolumn,english,showpacs,showkeys,preprintnumbers,floatfix,nofootinbib,11pt]{revtex4-2}
\usepackage{eurosym}
\usepackage{amssymb}
\usepackage{graphicx,color}
\usepackage{epsfig}
\usepackage{rotating}
\usepackage[T1]{fontenc}
\usepackage{latexsym,epsfig}
\usepackage[latin9]{inputenc}
\usepackage{amsfonts}
\usepackage{dcolumn}
\usepackage{bm}
\usepackage{babel}
\usepackage{hyperref}
\usepackage{amsmath,mathrsfs}
\pdfoutput=1
\begin{document}

\title{Efficient Computational method using random matrices describing critical thermodynamics}
\author{Roberto da Silva$^{1}$, Eliseu Venites$^{1}$, Sandra D. Prado$^{1}$,
J. R. Drugowich de Felicio$^{2}$}

\address{1 - Instituto de F{\'i}sica, Universidade Federal do Rio Grande do Sul,
Av. Bento Gon{\c{c}}alves, 9500 - CEP 91501-970, Porto Alegre, Rio Grande do
Sul, Brazil\\
2 - Departamento de F\'{\i}sica, Faculdade de Filosofia Ci\^{e}ncias e Letras de
Ribeir\~{a}o Preto, Universidade de S\~{a}o Paulo, 
Av. dos Bandeirantes 3900, Ribeir\~{a}o Preto, S\~{a}o Paulo, Brazil}


\begin{abstract}
Our research highlights the effectiveness of utilizing matrices akin to Wishart matrices, 
derived from magnetization time series data under specific dynamics, to elucidate phase 
transitions and critical phenomena in the Q-state Potts model. By employing appropriate 
statistical methods, we not only discern second-order transitions but also differentiate weaker 
first-order transitions through careful analysis of the density of eigenvalues and their fluctuations. 
Furthermore, we investigate the method's sensitivity to stronger first-order transition points. 
Importantly, we establish a robust correlation between the system's actual thermodynamics 
and the spectral thermodynamics encapsulated within the eigenvalues. Our findings are further 
substantiated by correlation histograms of the time series data, revealing insightful patterns. 
Expanding upon our core findings, we present a didactic analysis that draws parallels between 
the spectral properties of criticality in a spin system and matrices intentionally imbued with 
correlations (a toy model). Within this framework, we observe a universal behavior characterized 
by the distribution of eigenvalues into two distinct groups, separated by a gap dependent on the 
level of correlation, influenced by temperature-induced changes in the spin system.
\end{abstract}

\maketitle

\section{Introduction}

Fluctuation phenomena form the bedrock of understanding thermodynamics, 
encompassing both its processes and postulates \cite{Callen,MarioTermo,Salinas}. 
While Wigner's groundbreaking research, particularly his exploration of the intricate spectra of 
heavy nuclei mirrored in the eigenvalues of random matrices, did not primarily 
delve into thermodynamics \cite{Wigner,Wignerb,Wigner2,Mehta}, Dyson later unveiled 
a profound connection. Dyson elucidated the significant link between the joint 
eigenvalues distribution of random matrices and the Boltzmann weight of a static 
Coulomb gas, characterized by a logarithmic repulsion term between the charges 
and a harmonic attraction term \cite{Dyson,Dyson2,Dyson3,Mehta}.

Despite this progress, the direct correlation between the fluctuation phenomena 
of specific random matrices derived from natural systems and the thermodynamics 
of those systems remained elusive. Among numerous statistical concepts, correlation 
emerges as pivotal for understanding phase transitions and critical phenomena within 
Thermostatistics. Notably, Wishart \cite{Wishart}, several decades ahead of Wigner 
and Dyson, delved into correlated time series. He introduced the Wishart ensemble, 
which encompasses random correlation matrices, diverging from the Gaussian or Unitary 
ensembles.

The research presents an innovative approach utilizing Wishart matrices derived from 
magnetization time series to analyze the thermodynamic properties of eigenvalues linked 
to the Potts model. The findings indicate that this method adeptly captures both first-order 
phase transitions and critical phenomena, primarily by examining the dispersion of eigenvalues. 
Moreover, supplementary analyses demonstrate, through both analytical and computational means, 
the distribution of eigenvalue spectra within artificially correlated Wishart matrices, revealing discernible 
groups with a gap influenced by correlation strength, a phenomenon presented in a manner similar to the 
spectra of the Potts model. Significantly, these results highlight the versatility of the proposed methodology, 
extending its applicability beyond critical points to encompass first-order transition points as well. However, 
for the stronger first-order points, the method is not obligated to work, but even so, it seems to respond 
satisfactorily.

For that, it is essential to establish a brief review of the literature that
led to building our idea and its solution. Here we define the main object
for our analysis -- the matrix element $a_{ij} $ that denotes the amount of
the $j$-th time series at the $i$-th time step of a system with $N$
different time series. Here $i=1,...,M$, and $j=1,...,N$. So the matrix $A$
is $M\times N$. In order to analyze spectral properties, an interesting
alternative is to consider not $A$ but the square matrix $N\times $ $N$:

\begin{equation*}
G=\frac{1}{M}A^{T}A\ ,
\end{equation*}%
such that $G_{ij}=\frac{1}{M}\sum_{k=1}^{M}a_{ki}a_{kj}$, which is the
covariance matrix. It's crucial to note that $M\geq N$. In this context,
Wishart matrices are random matrix models that describe universal aspects of
covariance matrices.

At this juncture, rather than working directly with $A$, it becomes more
advantageous to work with the matrix $\Lambda $, defined in terms of
standard variables: 
\begin{equation}
\Lambda _{ij}=\frac{a_{ij}-\left\langle a_{j}\right\rangle }{\sigma _{a_{j}}}%
\text{,}
\end{equation}%
with $\sigma _{a_{j}}\approx \sqrt{\left\langle a_{j}^{2}\right\rangle
-\left\langle a_{j}\right\rangle ^{2}}$, where: $\left\langle
a_{j}^{k}\right\rangle =\frac{1}{M}\sum_{i=1}^{M}a_{ij}^{k}$ represents the $k$-- th
moment of $j$-- column. Thereby if $%
\mathcal{G} =\frac{1}{M}$\ $\Lambda ^{T}\Lambda $, we have: 
\begin{equation}
\begin{array}{lll}
\mathcal{G} _{ij} & = & \frac{1}{M}\sum_{k=1}^{M}\frac{a_{ki}-\left\langle
a_{i}\right\rangle }{\sqrt{\left\langle a_{i}^{2}\right\rangle -\left\langle
a_{i}\right\rangle ^{2}}}\frac{a_{kj}-\left\langle a_{j}\right\rangle }{%
\sqrt{\left\langle a_{j}^{2}\right\rangle -\left\langle a_{j}\right\rangle
^{2}}} \\ 
&  &  \\ 
& = & \frac{\left\langle a_{i}a_{j}\right\rangle -\left\langle
a_{i}\right\rangle \left\langle a_{j}\right\rangle }{\sigma _{a_{i}}\sigma
_{a_{j}}}\text{,}%
\end{array}%
\end{equation}%
where $\left\langle a_{i}a_{j}\right\rangle =\frac{1}{M}%
\sum_{k=1}^{M}a_{ki}a_{kj}$, i.e., $\mathcal{G} \ $is the correlation matrix.
Analytically, if $\Lambda _{ij}$\ ($a_{ij})$\ are independent random
variables, thus we have the so called Wishart orthogonal ensemble (WOE). On
the other hand, many authors consider to understand the role of correlations
and in that case, it is interesting to consider the one-sided correlated
Wishart orthogonal ensemble (CWOE) \cite{Seligman3,Novaes} and in this case $\Lambda ^{T}=\Omega
^{1/2}B^{T}$\ where $\Omega $\ is a real symmetric positive defnite
non-random $N\times N$\ matrix that accounts for the correlations in time
series (columns) of data matrix $\Lambda $\ and $B$\ is composed by
independent random variables with average 0 and variance 1. In this case 
\begin{equation*}
\mathcal{G} =\frac{1}{M}\Lambda ^{T}\Lambda =\frac{1}{M}\Omega ^{1/2}B^{T}B\ \Omega
^{1/2}
\end{equation*}%
which implies: 
\begin{equation}
Tr(\frac{1}{M}B^{T}B)=Tr(\frac{1}{M}\Omega ^{-1}\Lambda ^{T}\Lambda
)=Tr(\Omega ^{-1}\mathcal{G} )  \label{Eq:traco}
\end{equation}%
\ 

Now we know that the joint probability density of the matrix elements of $%
B^{T}$\ can be written considering gaussian distribution (no loss of
generality) for these entries: 
\begin{equation}
\begin{array}{lll}
P(B_{11},B_{12},...,B_{MN})dB & = & \prod\limits_{i=1}^{N}\prod%
\limits_{j=1}^{M}P(B_{ji})dB_{ji} \\ 
&  &  \\ 
& = & \frac{\exp (-\frac{1}{2}\sum_{i=1}^{N}\sum_{j=1}^{M}B_{ji}^{\ 2})}{%
(2\pi )^{NM/2}}dB \\ 
&  &  \\ 
& = & \frac{\exp (-\frac{1}{2}\sum_{i=1}^{N}[B^{T}B]_{ii})}{(2\pi )^{NM/2}}dB
\\ 
&  &  \\ 
& = & \frac{\exp \left[ -\frac{1}{2}Tr(B^{T}B)\right] }{(2\pi )^{NM/2}}dB%
\end{array}
\label{Eq:ProbB}
\end{equation}%
where $dB=\prod\limits_{i=1}^{N}\prod\limits_{j=1}^{M}dB_{ji}$.

Based on the equations \ref{Eq:traco} and \ref{Eq:ProbB} to write the joint
probability density for elements of $\mathcal{G} $\ supposes to calculate a
Jacobian from $M\times N$\ variables: $\left\{
B_{11},B_{12},...,B_{MN}\right\} \ $to $N\times N$\ variables: $\left\{ \mathcal{G}
_{11},\mathcal{G} _{12},...,\mathcal{G} _{NN}\right\} $\ which was performed by Wishart 
\cite{Wishart} and revisited by other authors (for example: \cite%
{Wilks,Muirhead}) which is given by:%
\begin{equation}
\begin{array}{ccc}
P(\mathcal{G} _{11},...,\mathcal{G} _{NN})d\mathcal{G} & = & K_{N}\det (\Omega )^{-\frac{M}{2}%
}\det (\mathcal{G} )^{\frac{(M-N-1)}{2}} \\ 
&  &  \\ 
&  & \times \exp \left[ -\frac{M}{2}tr(\Omega ^{-1}\mathcal{G} )\right] d\mathcal{G}%
\end{array}%
\end{equation}%
where $d\mathcal{G} =\prod\limits_{i=1}^{N}\prod\limits_{j>i}^{N}d\mathcal{G} _{ij}\times
\prod\limits_{i=1}^{N}d\mathcal{G} _{ii}$.

From that distribution we can obtain the jointly distribution of eigenvalues
of $\mathcal{G} $. In the particular case of $\Omega =1$, i.e., $\Lambda =B$, the
jointly distribution of eigenvalues is described by the Boltzmann weight 
\cite{GuhrPR,Seligman3}: 
\begin{equation}
\begin{array}{ccc}
P(\lambda _{1},...,\lambda _{N}) & = & C_{N}\exp \left[ -\frac{M}{2}%
\sum_{i=1}^{N}\lambda _{i}+\frac{M-N-1}{2}\sum_{i=1}^{N}\ln \lambda
_{i}\right. \\ 
&  &  \\ 
&  & \left. +\sum_{i<j}\ln \left\vert \lambda _{i}-\lambda _{j}\right\vert 
\right]%
\end{array}%
\end{equation}%
where $C_{N}^{-1}=\int_{0}^{\infty }...\int_{0}^{\infty }d\lambda
_{1}...d\lambda _{N}\exp [-H(\lambda _{1}...\lambda _{N})]$, corresponding
to the Hamiltonian:%
\begin{equation}
\mathcal{H}(\lambda _{1}...\lambda _{N})=\frac{M}{2}\sum_{i=1}^{N}\lambda
_{i}-\frac{(M-N-1)}{2}\sum_{i=1}^{N}\ln \lambda _{i}-\sum_{i<j}\ln
\left\vert \lambda _{i}-\lambda _{j}\right\vert
\end{equation}%
at temperature $\beta ^{-1}=1$. The last term is a logarithmic repulsion
exactly as the standard Wigner/Dyson ensembles.

Marchenko and Pastur \cite{Marcenko} showed that the density of eigenvalues
here written as: 
\begin{equation}
\sigma (\lambda )=\int_{0}^{\infty }...\int_{0}^{\infty }P(\lambda ,\lambda
_{2},...,\lambda _{N})d\lambda _{2}...d\lambda _{N}
\end{equation}%
of the matrix $\Phi =\frac{1}{M}\Lambda ^{T}\Lambda $, for the particular
case $\Omega =1$, follows the known Marcenko-Pastur distribution and not the
standard semi-circle law, described as:

\begin{equation}
\sigma (\lambda )=\left\{ 
\begin{array}{l}
\dfrac{M}{2\pi N}\dfrac{\sqrt{(\lambda -\lambda _{-})(\lambda _{+}-\lambda )}%
}{\lambda }\ \text{if\ }\lambda _{-}\leq \lambda \leq \lambda _{+} \\ 
\\ 
0\ \ \ \ \text{otherwise}%
\end{array}%
\right.  \label{Eq:MP}
\end{equation}%
where 
\begin{equation}
\lambda _{\pm }=1+\frac{N}{M}\pm 2\sqrt{\frac{N}{M}}.
\end{equation}

It is imperative to underscore that, within the framework of the Gaussian
random matrix model, a fundamental assumption is the existence of the first
four moments. This assumption implies that the presence of Levy tails in the
time series should not exhibit excessive heaviness. The presence of the
third and fourth moments holds particular significance as it enables control
over the statistical characteristics of the largest eigenvalues, ensuring
their alignment with the Gaussian random matrix model. This phenomenon is
extensively documented in the context of convergence to the Marchenko-Pastur
law, as exemplified in the work of Gotze and Tikhomirov \cite{Gotze}. It is
worth noting that the requirement for the fourth moment is not a mere
technical constraint; it plays a pivotal role in achieving convergence,
especially at the spectral edges.

However, the problem is highly complicated in the general 
case ($\Omega \neq 1$), and the spectra do not precisely follow 
a simple closed form. Closed forms were interestingly obtained by Guhr 
and collaborators using supersymmetric methods \cite{Guhr}. 
They also used similar approach to study the statistics
of the smallest eigenvalues \cite{Guhr2}. An even more detailed result is
presented in Sections III and IV of the work by Wirtz et al. \cite{Wirtz}.
Prior to that, the connection between random matrices and critical phenomena
had other important actors.

Some interesting contributions in the late 1980s and early 1990s 
(see, for example, \cite{Kostov1989,Cicuta1990}) explored the relationship 
between critical properties of statistical mechanics models in equilibrium and the 
spectral density of random matrices. It is also known that the largest eigenvalue of 
a complex Gaussian sample covariance matrix, as well as others, such as those from 
the Gaussian family, exhibit sharp phase transitions, which have been studied in the 
literature (see, for example, \cite{Baik,Georges}).

In an exciting application of random matrices the authors in \cite%
{Stanley,Stanleyb,Stanley2}, and simultaneously and independently in \cite%
{Bouchaud,Bouchaud2} using the results developed by Marcenko and Pastur \cite%
{Marcenko,Sengupta}, showed that deviations from the bulk of spectra of
random correlation matrices built with financial market assets are related
to genuine correlations from Stock Market. Emphasizing the significance of 
exploring the structure of covariance matrices in Principal Component 
Analysis is paramount, as demonstrated by seminal works such 
as \cite{Johnstone}.

Interestingly, some authors
investigated spectral properties of correlation matrices in near-equilibrium
phase transitions \cite{Vinayak2014}. In this case, they studied correlation
matrices of the $N=L^{2}$ spins of the Ising model in the two-dimensional
lattice under $\tau $ time steps of evolution to evidence of the power-law
spatial correlations at a phase transition display. Similarly, the authors
in \cite{Biswas2017} explored results in the steady-state for the
correlation matrix of the asymmetric simple exclusion process.

However, besides the body of results in this field, we are far from grasping
all the information about phase transitions in spin systems. They seem to be
inherent to the system even in its initial stages of evolution \cite%
{Zhengprimordial,Huse1989}. When suddenly placed at finite temperatures, a system
initially at high temperatures responds to the natural pre-existent natural
correlations in the interacting system. For instance, it can explain the
richness of the time evolution of magnetization \cite{Janssen,Janssen2}.

In this context, the most relevant theories consider that the critical
behavior of the statistical mechanics systems can be captured before the
system reaches the equilibrium, and scaling relations that consider initial
conditions were used to obtain critical exponents and parameters (see, for
example: \cite{Zhengprimordial,Huse1989,Janssen,Janssen2,Zheng,Zheng2,Zheng3}). These results motivated
several other approaches in different systems with and without defined
Hamiltonian, and mean-field systems (only to mention some relevant
contributions: \cite{Pleimling,Pleimling2,Pleimling3,Pleimling4,Pleimling5,Pleimling6}).

Recently, one of the authors of this current work showed that random
matrices could enlighten critical phenomena by showing that the density of
eigenvalues strongly responds to the criticality of the system in a simple
model as the two-dimensional Ising model \cite{RMT2023}. Motivated by this
result, the question which arises is if the phase transitions are imprinted
into the properties of random matrices simulated via Monte Carlo (MC) from
time evolutions far from thermalization. Moreover, another question to be
answered is whether such a method works only for critical points or can also
be extended for first-order transition points. If extendable, are there
fundamental or universal aspects that go beyond spin models and are also
characterized by the correlations existing in time series, not only in the
magnetization time series?

In this paper, we give a positive answer to this question and supply the
fundamentals to understand how the eigenvalues density of random matrices
governs the phase transition points of the Potts model. We show that the
method describes very well the critical (second-order) points of the Potts
model which was its initial proposal for the Ising model in both short and
long-range mean field Ising model \cite{RMT2023,RMT-2}, but surprisingly the
method works very well for the a weak first order point $Q=5$. For $Q\geq 7$%
\ (stronger first order points) the method seems to respond for amounts
based on upper fluctuations of eigenvalues but it presents deviations for
the expect value $\left\langle \lambda \right\rangle $. It is important to
note, to the best of our knowledge, that methods for second and first order
are applied distinctly in the literature, and the method here presented has
no obligation to work for strong first order which will deserves atention in
other alternative study.

In addition, we show how the artificially correlated system's spectra
explain the existence of two groups of eigenvalues separated by a gap that
depends on the strength of the correlation. This gap dependence on the
correlation is the key to understanding what ocurrs for the critical and
first-order phenomena.

We thus provide an affirmative response to this pivotal question and lay the
foundational groundwork for comprehending how the eigenvalue density of
random matrices governs the phase transition points of the Potts model. Our
method excels in accurately characterizing the critical (second-order)
points of the Potts model, originally proposed for the Ising model, both in
the short and long-range mean field Ising model, as documented in \cite%
{RMT2023,RMT-2}. Remarkably, our method also proves effective in capturing a
weak first-order point at $Q=5$.

For cases where $Q\geq 7$, corresponding to stronger first-order phase
transitions, our method exhibits sensitivity to the dispersion of
eigenvalues and yields insights from this metric. However, it exhibits
deviations in the expectation value $\left\langle \lambda \right\rangle $.
It is worth emphasizing that, to the best of our knowledge, methods for
second-order and first-order phase transitions are typically distinct in the
literature. Our method, as presented here, is not inherently designed to
handle strong first-order transitions, which warrant further exploration in
alternative studies.

Additionally, we elucidate how the spectra of artificially correlated
systems shed light on the existence of two distinct groups of eigenvalues,
separated by a gap whose magnitude is contingent upon the strength of the
correlation. This gap's dependence on the correlation strength holds the key
to comprehending the behaviors associated with critical and first-order
phenomena.

We first performed preparatory results about the spectral properties of
random matrices built from correlated time series pair-by-pair, considering
a simple way to fix the correlation. We then explored the emergence of the
gap of eigenvalues and how this influences the fluctuation spectra from
simple analytical cases and after with MC simulations. Such a study is
presented in the next section (Sec. \ref{Sec:Correlations_preparatory_study}%
).

In section \ref{Sec:Potts}, we present the details about how to build the
Wishart matrices of the time series of magnetization for the Potts model.
Since we observed how the correlations between time series could influence
the density of eigenvalues in Sec. \ref{Sec:Correlations_preparatory_study},
we then performed simulations for the Potts model from $q=3$ up to $q=10$ by
showing how the spectra of random matrices built from magnetization time
series can precisely indicate the transition points and, most importantly,
independently from the transition order since we migrate from critical
points passing by weak first-order points until strong first-order
transition. We present our main results in section \ref{Sec:Results}.

To support our results even more, we analyzed the histograms of correlations
between the different evolutions, i.e., the histogram of elements of Wishart
matrices, and observed how such histograms are linked to the density of
eigenvalues.

The gap in the distribution of eigenvalues seems to govern the existence of
phase transitions but not in a straightforward way, and this can be observed
in even simpler situations, i.e., not for a physical system, but for a
system artificially built, as suggested by the spectra studied in the
preparatory section (Sec. \ref{Sec:Correlations_preparatory_study})

The main contribution of this work is strongly based on the proposal of a
spectral thermodynamics built from Wishart matrices that can reflect the
real thermodynamics of spin systems.

\section{Correlations and Wishart matrices: a preparatory study}

\label{Sec:Correlations_preparatory_study}

In this section, we show the role played by correlations in the emergence of
a gap in the distribution of the eigenvalues of Wishart matrices is the
eigenvalues of Wishart matrices. For that, we build Wishart matrices by
artificially correlating columns of the matrix $A$ by pairs. A simple way to
correlate two random variables is the one used by us in two previous papers.
One to correlate phases of waves to observe the emerging of rogue waves \cite%
{Silvacorr1} and a second to spatially correlate points in the traveling
salesman problem to analyze the loss of performance of simulated annealing
from two to one dimension \cite{Silvacorr2}. Consider two independent and
identically distributed (i.i.d.) random variables $\varphi _{1}$ and $%
\varphi _{2}$ with $\left\langle \varphi _{1}\right\rangle =\left\langle
\varphi _{2}\right\rangle =0$. Then, two $\rho -$correlated variables can be
obtained by making $\phi _{1}=$ $\varphi _{1}\sin \theta +\varphi _{2}\cos
\theta $ and $\phi _{2}=$ $\varphi _{1}\cos \theta +\varphi _{2}\sin \theta $%
, if:%
\begin{equation}
\rho =\frac{\left\langle \left( \ \phi _{1}-\left\langle \phi
_{1}\right\rangle \right) \left( \ \phi _{2}-\left\langle \phi
_{2}\right\rangle \right) \right\rangle }{\sqrt{\left\langle \left( \Delta
\phi _{1}\right) ^{2}\right\rangle \left\langle \left( \Delta \phi
_{2}\right) ^{2}\right\rangle }}=\sin 2\theta .
\end{equation}%
where we assume the existence of the second moments of $\phi _{1}$\ and $%
\phi _{2}$, i.e., $0<\left\langle \left( \Delta \phi _{1}\right)
^{2}\right\rangle <\infty $\ and $0<\left\langle \left( \Delta \phi
_{2}\right) ^{2}\right\rangle <\infty $, with $\left\langle \left( \Delta
\phi _{1}\right) ^{2}\right\rangle =\left\langle \phi _{1}^{2}\right\rangle
-\left\langle \phi _{1}\right\rangle ^{2}$, and $\left\langle \left( \Delta
\phi _{2}\right) ^{2}\right\rangle =\left\langle \phi _{2}^{2}\right\rangle
-\left\langle \phi _{2}\right\rangle ^{2}$.

In order to understand the role played by $\rho $ on the spectra of $\mathcal{G}$, we
build matrices $A$ where each pair of adjacent columns are composed by two
vectors $\mathbf{a}_{i}=\left[ a_{i,1}...a_{i,M}\right] ^{t}$ and $\mathbf{a}%
_{i+1}=\left[ a_{i+1,1}...a_{i+1,M}\right] ^{t}$, such that: $%
a_{k,i}=\varphi _{k,i}^{(1)}\sin \theta +\varphi _{k,i}^{(2)}\cos \theta $,
while $a_{k,i+1}=\varphi _{k.i}^{(1)}\cos \theta +\varphi _{k,i}^{(2)}\sin
\theta $ (recall that this is not a rotation). It follows straightforward
from the fact that $\varphi _{i,k}^{(1)}$ and $\varphi _{i,k}^{(2)}$ are $%
\rho -$correlated random variables that $a_{k,i} $ and $a_{k,i+1}$ are also
two $\rho -$correlated random variables with average zero and variance one.

By choosing $\varphi _{i,k}^{(1)}$, $\varphi _{i,k}^{(2)}$, $k=1,...,M$ ,
random variables $N[0,1]$ we build matrices where some columns are
correlated. For $N=2$, we certainly guarantee that the correlation between
the columns is $\rho $. Particularly for $M=2$, the elements are:%
\begin{equation}
\mathcal{G}_{11}=\varphi _{1,1}^{(1)2}\sin ^{2}\theta +\varphi _{1,1}^{(2)2}\cos
^{2}\theta +2\varphi _{1,1}^{(1)}\varphi _{1,1}^{(2)}\sin \theta \cos \theta
\end{equation}%
such that $\left\langle \mathcal{G}_{11}\right\rangle =1$, since $\left\langle \varphi
_{1,1}^{(1)2}\right\rangle =\left\langle \varphi _{1,1}^{(2)2}\right\rangle
=1$ and $\left\langle \varphi _{1,1}^{(1)}\varphi _{1,1}^{(2)}\right\rangle
=\left\langle \varphi _{1,1}^{(1)}\right\rangle \left\langle \varphi
_{1,1}^{(2)}\right\rangle =0$. In that same way, we can conclude that: $%
\left\langle \mathcal{G}_{12}\right\rangle =\left\langle \mathcal{G}_{21}\right\rangle =2\sin
\theta \cos \theta =\rho $ and $\left\langle \mathcal{G}_{22}\right\rangle =1$. It
follows that the eigenvalues of $\left\langle \mathcal{G}\right\rangle $ in this case
are $\lambda _{\pm }=1\pm \rho $.

\begin{figure*}[tbp]
\begin{center}
\includegraphics[width=1.0\columnwidth]{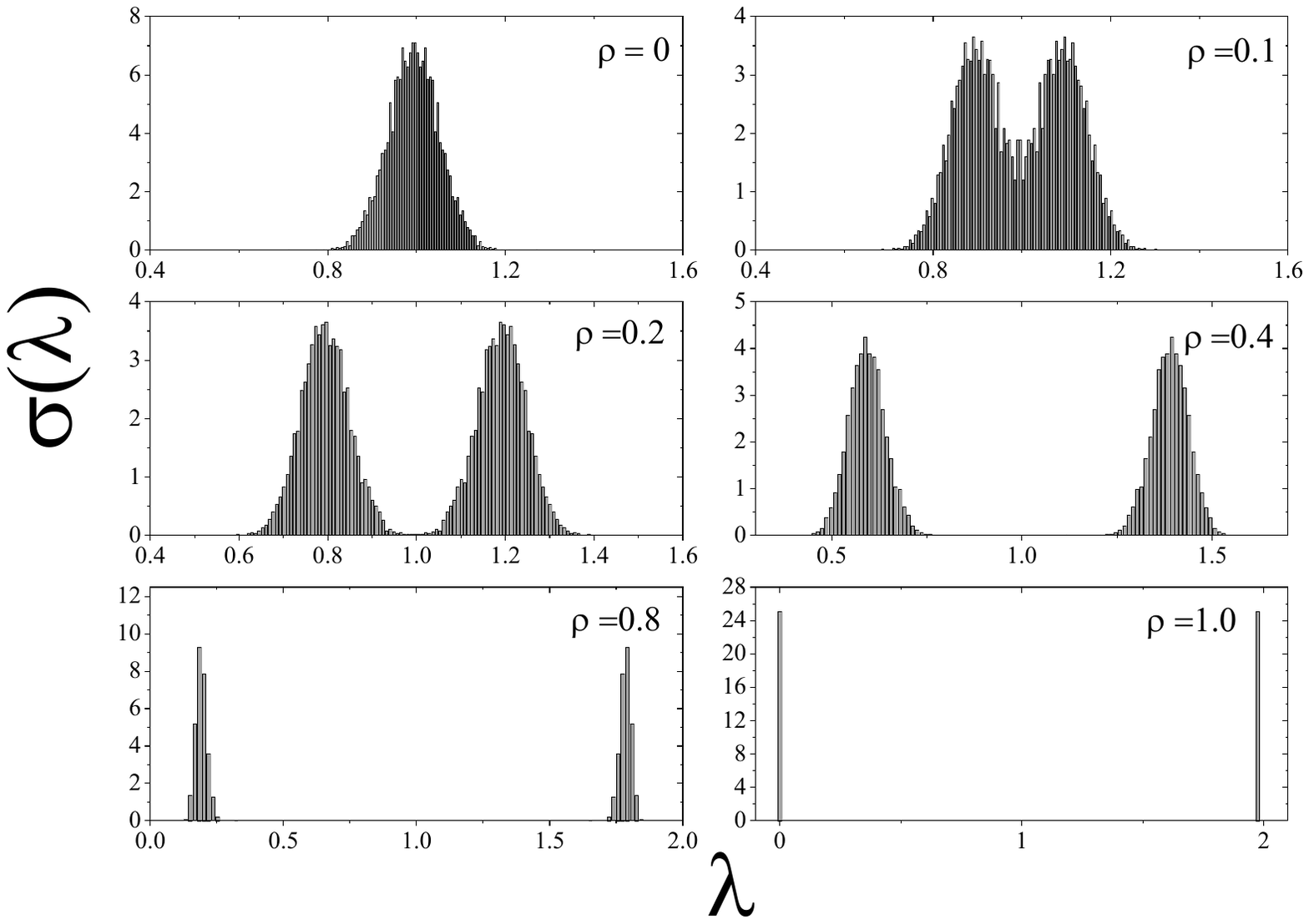}
\end{center}
\caption{Density of eigenvalues $\mathcal{G}$ for different values of $\protect\rho $
considering an ensemble of $N_{run}=4000\ $ different matrices. The pairwise
columns of $A$ ($\mathcal{G}=\frac{1}{M}\Lambda^{t}\Lambda$) are $\protect\rho $-correlated
according to our procedure. We used $N=2$ and $M=300$. }
\label{Fig:Density_of_eigenvalues_N=2_M=300}
\end{figure*}

For an arbitrary $M$, one has a block-diagonal matrix written as 
\begin{equation}
\left\langle \mathcal{G}\right\rangle =\left( 
\begin{array}{cc}
1 & \rho \\ 
\rho & 1%
\end{array}%
\right) \otimes \mathbf{I}_{N/2\times N/2}
\end{equation}%
and the eigenvalues of $\left\langle \mathcal{G}\right\rangle $ are similarly $\lambda
_{\pm }=1\pm \rho $ with multiplicity $N/2$. However, the situation is quite
different in the case of an ensemble of matrices $G$ corresponding to the
different values of $\varphi _{i,k}^{(1)}$ and $\varphi _{i,k}^{(2)}$ for $%
k=1,...,M$ \ and $i=1,...,N/2$. By instance, for $M=300$ and $N=2$, with $%
N_{run}=4000$ different matrices $\mathcal{G}$, the density of eigenvalues $\sigma
(\lambda )$ is two-peaked. The peaks are in $1-\rho $ and other in $1+\rho $
as illustrated in Fig. \ref{Fig:Density_of_eigenvalues_N=2_M=300}, in
agreement with the result obtained for $\left\langle \mathcal{G}\right\rangle $ for
the case of $M=2$.

However, this is not true for $N=100$. In this case, for $\rho =0$ and $N$
large, the density of eigenvalues must behave as a Marchenko-Pastur law and
not only two eigenvalues with multiplicity $N/2$. Then, it is important to
check what happens for $\rho >0$ and $N$ large.

\begin{figure*}[tbp]
\begin{center}
\includegraphics[width=1.0\columnwidth]{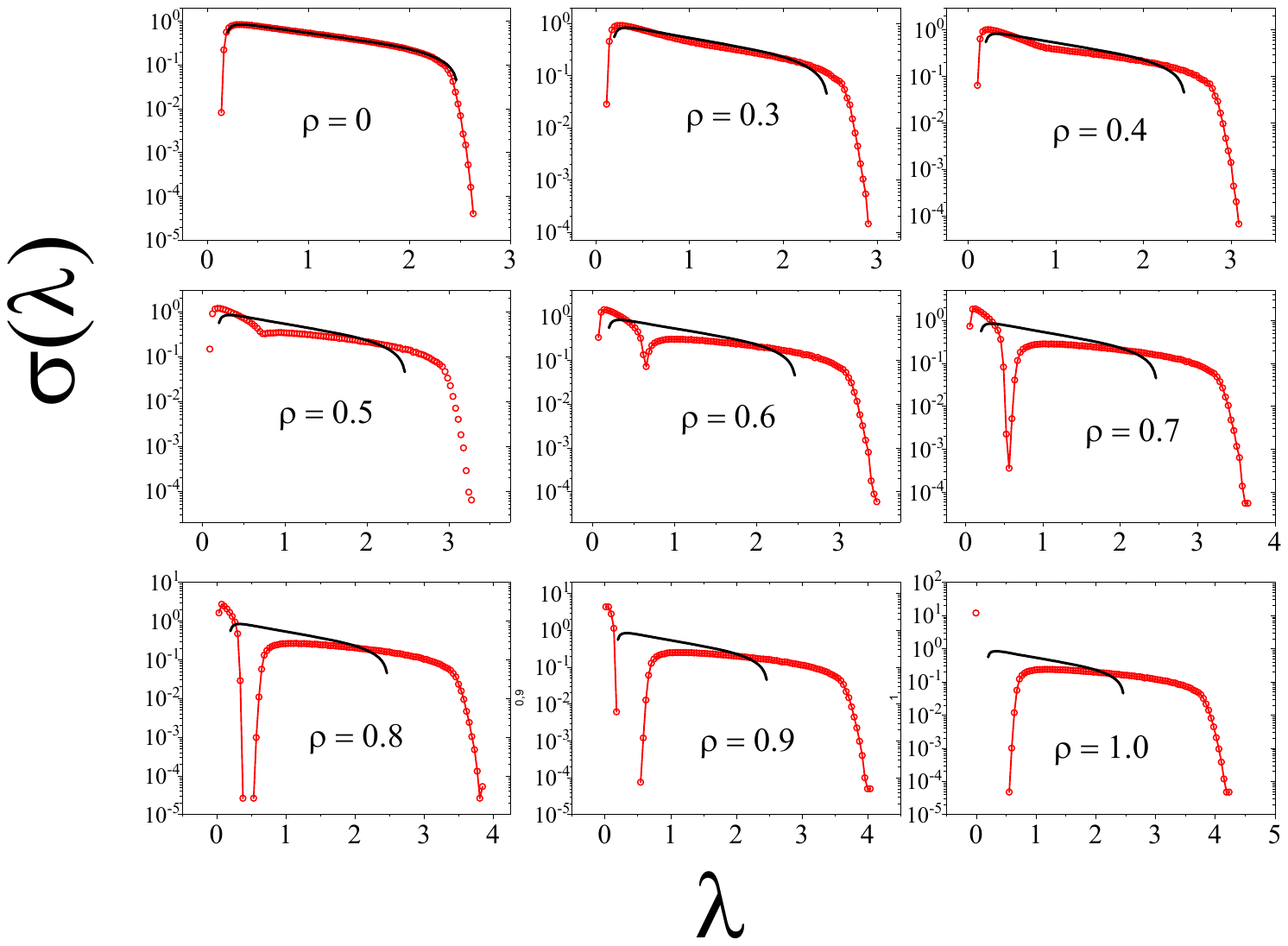}
\end{center}
\caption{Density of eigenvalues considering an ensemble of $N_{run}=4000\ $%
different matrices $\mathcal{G}$ for different values of $\protect\rho $. The pairwise
columns of $A$ ($\mathcal{G}=\frac{1}{M}\Lambda^{t}\Lambda$) are $\protect\rho $-correlated
according to our procedure. We used $N=100$ and $M=300$. The continuous
curve corresponds to the Marchenko-Pastur density of eigenvalues.}
\label{Fig:Density_of_eigenvalues_N=100_M=300}
\end{figure*}

Fig. \ref{Fig:Density_of_eigenvalues_N=100_M=300} shows that as $\rho $
increases, the density of eigenvalues becomes distinguished from the
Marchenko-Pastur prediction, as expected. We also observe a singularity that
becomes more pronounced as $\rho$ increases. This discontinuity gives origin
to a gap between $\rho =0.7$ and $\rho =0.8$. This gap enlarges as $\rho$
increases until one of the branches disappears ($0.9 <\rho <1.0$). Here, the
effective correlation is not exactly $\rho $ since when crossing columns of
no adjacent columns, the correlation is zero on average (or 1 between the
same columns). On the other hand, it is a fact that when $\rho $ increases,
the effective correlation also increases.

To examine the influence of correlation $\rho $\ on eigenvalue fluctuations,
we plotted $\left\langle \lambda \right\rangle $\ and $\zeta =\frac{d}{d\rho 
}\left\langle (\Delta \lambda )^{2}\right\rangle $\ as functions of $\rho $,
where $\left\langle (\Delta \lambda )^{2}\right\rangle =\left\langle \lambda
^{2}\right\rangle -\left\langle \lambda \right\rangle ^{2}$\ is the
dispersion of the eigenvalues. Notably, we observed that at $\rho \approx
0.9 $, there was a maximum value of $\left\langle \lambda \right\rangle $\
and a corresponding minimum value of $\zeta $\ (refer to Fig. \ref%
{Fig:Fluctuations_as_function_of_rho}). In Fig. \ref%
{Fig:Density_of_eigenvalues_N=100_M=300}, we can observe an interesting
competition in the evolution of eigenvalue density: as $\rho $\ increases,
the singularity previously mentioned causes the density of eigenvalues near
the origin ($\lambda \approx 0$) to increase while the bulk of eigenvalues
on the right side of the singularity slightly shifts to the right. The
former phenomenon appears to have a stronger effect than the latter as $\rho 
$\ increases, leading to a decrease in the average eigenvalue until $\rho
\approx 0.9$. However, an inversion is observed for $\rho >0.9$, where the
average eigenvalue increases due to a stagnation of null eigenvalues caused
by the high correlation and the right shift being more preponderant. The
derivative of the dispersion of eigenvalues seems to follow this trend,
decreasing for $\rho >0.9$\ after an initial increase leading up to this
value.

It's worth noting that Figure \ref{Fig:Density_of_eigenvalues_N=100_M=300}
indicates a superposition of two Marchenko-Pastur laws, which becomes
evident once two distinct spectral bulks have been identified, as observed
by Wirtz, Kieburg, and Guhr \cite{Wirtz}. 
\begin{figure}[tbp]
\begin{center}
\includegraphics[width=0.8\columnwidth]{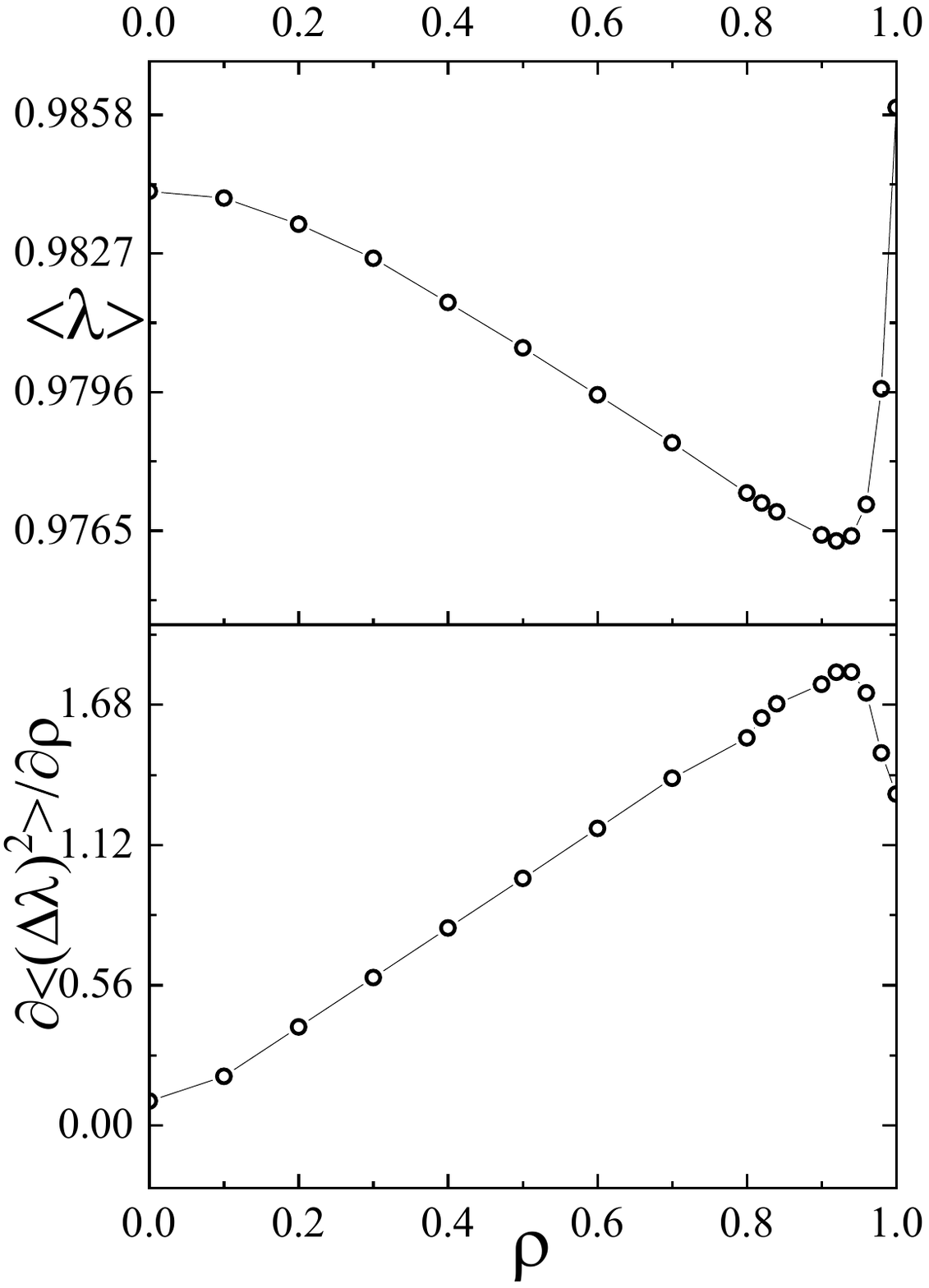}
\end{center}
\caption{Eigenvalue average and derivative of eigenvalue dispersion as a
function of the $\protect\rho $. Extremals are observed for $\protect\rho %
_{c}\approx 0.9$. }
\label{Fig:Fluctuations_as_function_of_rho}
\end{figure}

This behavior implies that eigenvalues are correlated, and their dependence
on these correlations can be quantified by examining their fluctuations. It
is worth noting that this pedagogical study bears resemblances to
significant prior findings in the literature, specifically the spiked
(Wishart) random matrix ensemble, as discussed in references \cite%
{Bloemendal,Bloemendal2,Mo,Mo2}. Furthermore, it is advisable to verify the
results presented here by comparing them to the analytical outcomes in \cite%
{Wirtz}.

Using such motivation, we will consider now, Wishart matrices built from
time series of magnetization of the Potts model for an arbitrary number of
states $Q$.

We will show that we can identify phase transitions using the fluctuations
of eigenvalues of such matrices and that in this particular case, the
minimum of average eigenvalue occurs at the proximity of the transition
point, as well as an inflection point in the variance of eigenvalue.

Still, we also illustrate that our results are supported by monitoring the
histograms of correlation at different temperatures of the system.

\section{Potts model and Wishart matrices}

\label{Sec:Potts}

The Hamiltonian of the $d$-dimensional $Q-$state Potts model (see for
example \cite{Wu}) can be written as 
\begin{equation}
\beta \mathcal{H}_{\text{Potts}}=-K\sum_{\left\langle i,j\right\rangle }\
\delta _{s_{i},s_{j}}  \label{Eq:Hamiltonian}
\end{equation}%
where $\beta =(k_{B}T)^{-1}$, with $k_{B}$ being the Boltzmann constant, $T$
is the system's temperature, and $K$ is the coupling coefficient. Here $%
\left\langle i,j\right\rangle $ indicates that the sum is taken over all
nearest neighbor sites, and $s_{i}=0,1,...,Q-1$. The system presents a phase
transition for $T_{C}={K_{C}}^{-1}=(\ln (1+\sqrt{Q}))^{-1}$ which is
continuous (second order) for $Q\leq 4$ and discontinuous (first order) for $%
Q\geq 5$.

In addition to the many equilibrium studies of the Potts model, some authors
explored its critical dynamics \cite{Silva2004,Forgacs}. A non-equilibrium
version of the model, driven out of equilibrium by coupling the spins to
heat baths at two different temperatures presents entropy production, as the
authors interestingly showed in \cite{Martynec}.

Thus, now, we define the main object for our current analysis, the
magnetization matrix element:

\begin{equation}
\mathcal{M}_{tj}=\frac{1}{L^{d}(q-1)}\sum_{k=1}^{L^d}(q\delta _{s_{k}(t,j),1}-1)
\label{Eq:Traditional_order_parameter}
\end{equation}%
where $s_{k}(t,j)$ denotes the state of $k-$th spin of the $j$th time series
at the $t$th MC step of a system with $L^{d}$ spins (in this work, for
simplicity, we used $d=2$, the minimal dimension to appear phase transition
in the Potts model). Here $t=1,...,N_{MC}$, and $j=1,...,N_{run}$. In this
same equation, $\delta _{\mu ,\nu }$ denotes the Kronecker delta symbol,
which assumes a value equal to 1 if $\mu =\nu $ and 0 otherwise. So the
magnetization matrix $\mathcal{M}$ is $N_{MC}\times N_{run}$. Similarly, we define: $%
\mathcal{M}_{tj}^{\ast }=\frac{\mathcal{M}_{tj}-\left\langle \mathcal{M}_{j}\right\rangle }{\sqrt{%
\left\langle \mathcal{M}_{j}^{2}\right\rangle -\left\langle \mathcal{M}_{j}\right\rangle ^{2}}}$%
, exactly as we previously described and thus we obtain the Wishart matrix $%
\mathcal{G}=\frac{1}{N_{MC}}\mathcal{M^{\ast }}^{T}\mathcal{M^{\ast }}$.

Here it is important to elucidate a pivotal calculation that shows what we
are exactly with these matrices with more details. Consider two distinct
time evolutions, denoted as $\mathcal{M}_{tk}$\ and $\mathcal{M}_{tl}$, $t=1,...,N_{MC}\ $are
two different time evolutions of the magnetization per spin. In this
scenario, we can write in a more general way: 

\begin{equation}
\mathcal{M}_{tk}=\frac{1}{L^2}\sum_{i=1}^{L^2}\sigma _{t,k,i}\text{,}  \label{Eq:mag_new}
\end{equation}%

where $\sigma _{t,k,i}$\ represents the value associated with the $i$-th spin
variable in the $k$-th evolution or run at time $t$. For example for the Potts model, $\sigma _{t,k,i}=\frac{1}{(q-1)}(q\delta_{s_{i}(t,k),1}-1)$

We can establish the correlation between these two time series through the
following definition:

\begin{equation}
\begin{array}{lll}
\left\langle \mathcal{M}_{k}\mathcal{M}_{l}\right\rangle _{t} & = & \frac{1}{N_{MC}}%
\sum_{t=1}^{N_{MC}}\mathcal{M}_{tk}\mathcal{M}_{tl} \\ 
&  &  \\ 
& = & \frac{1}{L^{4}N_{MC}}\sum_{t=1}^{N_{MC}}\left( \sum_{i=1}^{L^2}\sigma
_{t,k,i}\right) \left( \sum_{j=1}^{L^2}\sigma _{t,l,j}\right) \\ 
&  &  \\ 
& = & \frac{1}{L^{4}N_{MC}}\sum_{t=1}^{N_{MC}}\left( \sum_{i=1}^{L^2}\sigma
_{t,k,i}\sigma _{t,l,i}+\sum_{i\neq j=1}^{L^2}\sigma _{t,k,i}\sigma
_{t,l,j}\right) \\ 
&  &  \\ 
& = & \frac{1}{L^{4}}\sum_{i=1}^{L^2}\left[ \left( \frac{1}{N_{MC}}%
\sum_{t=1}^{N_{MC}}\sigma _{t,k,i}\sigma _{t,l,i}\right) +\sum_{i\neq
j=1}^{L^2}\left( \frac{1}{N_{MC}}\sum_{t=1}^{N_{MC}}\sigma _{t,k,i}\sigma
_{t,l,j}\right) \right] \\ 
&  &  \\ 
& = & \frac{1}{L^{4}}\sum_{i=1}^{L^2}\left\langle \sigma _{k,i}\sigma
_{l,i}\right\rangle _{t}+\frac{1}{L^{4}}\sum_{i\neq j=1}^{L^2}\left\langle
\sigma _{k,i}\sigma _{l,j}\right\rangle _{t}%
\end{array}
\label{Eq:correlation}
\end{equation}

Given that $\sum_{i=1}^{L^2}\left\langle \sigma _{k,i}\sigma
_{l,i}\right\rangle _{t}=O(L^2)$, and $\sum_{i\neq j=1}^{L^2}\left\langle \sigma
_{k,i}\sigma _{l,j}\right\rangle _{t}=O(L^{4})$, if we assume a moderate
strength of alignments among spins in both space and time we have: 
\begin{equation*}
\left\langle \mathcal{M}_{k}\mathcal{M}_{l}\right\rangle _{t}\approx \left\langle \frac{1}{L^{4}}%
\sum_{i\neq j=1}^{L^2}\sigma _{k,i}\sigma _{l,j}\right\rangle
_{t}=\left\langle \sigma _{k}\otimes \sigma _{l}\right\rangle _{t}\text{.}
\end{equation*}

When $T>T_{C}$, $\left\langle \mathcal{M}_{k}\right\rangle _{t}\approx 0$. This leads
to: $\left\langle \mathcal{M}_{k}\mathcal{M}_{l}\right\rangle _{t}-\left\langle
\mathcal{M}_{k}\right\rangle _{t}\left\langle \mathcal{M}_{l}\right\rangle _{t}\approx \frac{1}{%
L^{4}}\left\langle \sigma _{k}\otimes \sigma _{l}\right\rangle _{t}$, and we
can express the correlation coefficient (our matrix element of $\mathcal{G}$) as:%
\begin{equation*}
\begin{array}{lllll}
\mathcal{G}_{kl} & \approx & \frac{\left\langle \mathcal{M}_{k}\mathcal{M}_{l}\right\rangle _{t}}{\sqrt{%
\left\langle \mathcal{M}_{k}^{2}\right\rangle _{t}}\sqrt{\left\langle
\mathcal{M}_{l}^{2}\right\rangle _{t}}} &  &  \\ 
&  &  &  &  \\ 
& \approx & \frac{\left\langle \mathcal{M}_{k}\mathcal{M}_{l}\right\rangle _{t}}{\left\langle
\mathcal{M}_{k}^{2}\right\rangle _{t}} &  &  \\ 
&  &  &  &  \\ 
& = & \frac{\left\langle \sigma _{k}\otimes \sigma _{l}\right\rangle _{t}}{%
\left\langle \sigma _{k}\otimes \sigma _{k}\right\rangle _{t}} & \approx & 
\frac{\left\langle \sigma _{k}\otimes \sigma _{l}\right\rangle _{t}}{%
\left\langle \sigma _{l}\otimes \sigma _{l}\right\rangle _{t}}%
\end{array}%
\end{equation*}%
where $\sigma _{k}\equiv (\sigma _{k,1},...,\sigma _{k,N})$\ and $\sigma
_{l}\equiv (\sigma _{l,1},...,\sigma _{l,N})$. Thus $\mathcal{G}_{kl}$\ for $T>T_{C}$
is determined by: 
\begin{equation*}
\mathcal{G}_{kl}\approx \frac{\left\langle \sigma _{k}\otimes \sigma _{l}\right\rangle
_{t}}{\left\langle \sigma _{l}\otimes \sigma _{l}\right\rangle _{t}}
\end{equation*}

This metric assesses the relationship between the temporal averages of
spatial correlations in both inter and intra time series. By analyzing both
spatial and temporal dimensions, it presents an intriguing avenue for
investigating spin systems.

Undoubtedly, being $\mathcal{M}_{ij}$ the average magnetization (Eq. \ref%
{Eq:Traditional_order_parameter}) , then we expect that for $T>>T_{c}$ the
density of eigenvalues $\sigma ^{\exp }(\lambda )$\ obtained from
computational simulations approaches $\sigma (\lambda )$ in Eq. \ref{Eq:MP}.
The interesting question is what happens when $T\approx T_{C}$. More than
that, we will use the density $\sigma ^{\exp }(\lambda )$, the one obtained
from computer simulations, to obtain the critical parameter of spin models.
We will show that the results here go beyond the Ising model (already
studied with a similar approach \cite{RMT2023}) since they give an account
of the first-order points in addition to the critical ones.

\section{Main results}

\label{Sec:Results}

We performed MC simulations to obtain the eigenvalues of an ensemble of $%
N_{run}=4000$ different random matrices $\mathcal{G}$ obtained from $%
N_{sample}=1000$ time evolutions of $q-$state Potts model of the $N_{MC}=300$
first MC steps of evolution at temperature $T$. We start with a system at
infinite temperature (spins randomly chosen) and simulated according to
Metropolis algorithm for systems of linear dimension $L=128$ (large enough
here as discussed for the Ising model in \cite{RMT2023}).

Just like it was done in Section \ref{Sec:Correlations_preparatory_study},
the eigenvalues of the ensemble of matrices were grouped. The maximum and
minimum among all of the eigenvalues of the ensemble defined the interval
that was divided into a fixed number of bins: $N_{bins}=100$. Thus we
plotted the shape of the density of states for different temperatures.

We start our discussion by showing the density of states for $Q=3$ with
three different temperatures. They are illustrated in Fig. \ref%
{Fig:density_potts3}. The density changes its shape as temperature
increases. There is a gap between two groups of eigenvalues for temperature
below the critical temperature ($T=1/2T_{C}$). The gap vanishes above the
critical temperature ($T=1.05T_{C}$). More precisely, the gap is smaller as $%
T$ approaches $T_{C}$, and it vanishes for temperatures close to the
critical temperature as reported for the case $Q=2$ (see \cite{RMT2023}).

For high temperatures, we observe a better approximation for the
Marchenko-Pastur (red curve) in the plot corresponding to $T=6.5T_{C}$,
since the system becomes uncorrelated. However not by the tail that differs
on the Marchenko-Pastur law. This ocurrs because we consider the correlation
between the total magnetization not extracting the autocorrelation spin-spin
that generate this distinct tail in relation to expect law. It is important
to mention that it is not importance to the working of the method.

Recall that time evolutions start from initial conditions at infinite
temperatures, i.e. spins are randomly chosen with the same probability.

\begin{figure}[tbp]
\begin{center}
\includegraphics[width=0.8\columnwidth]{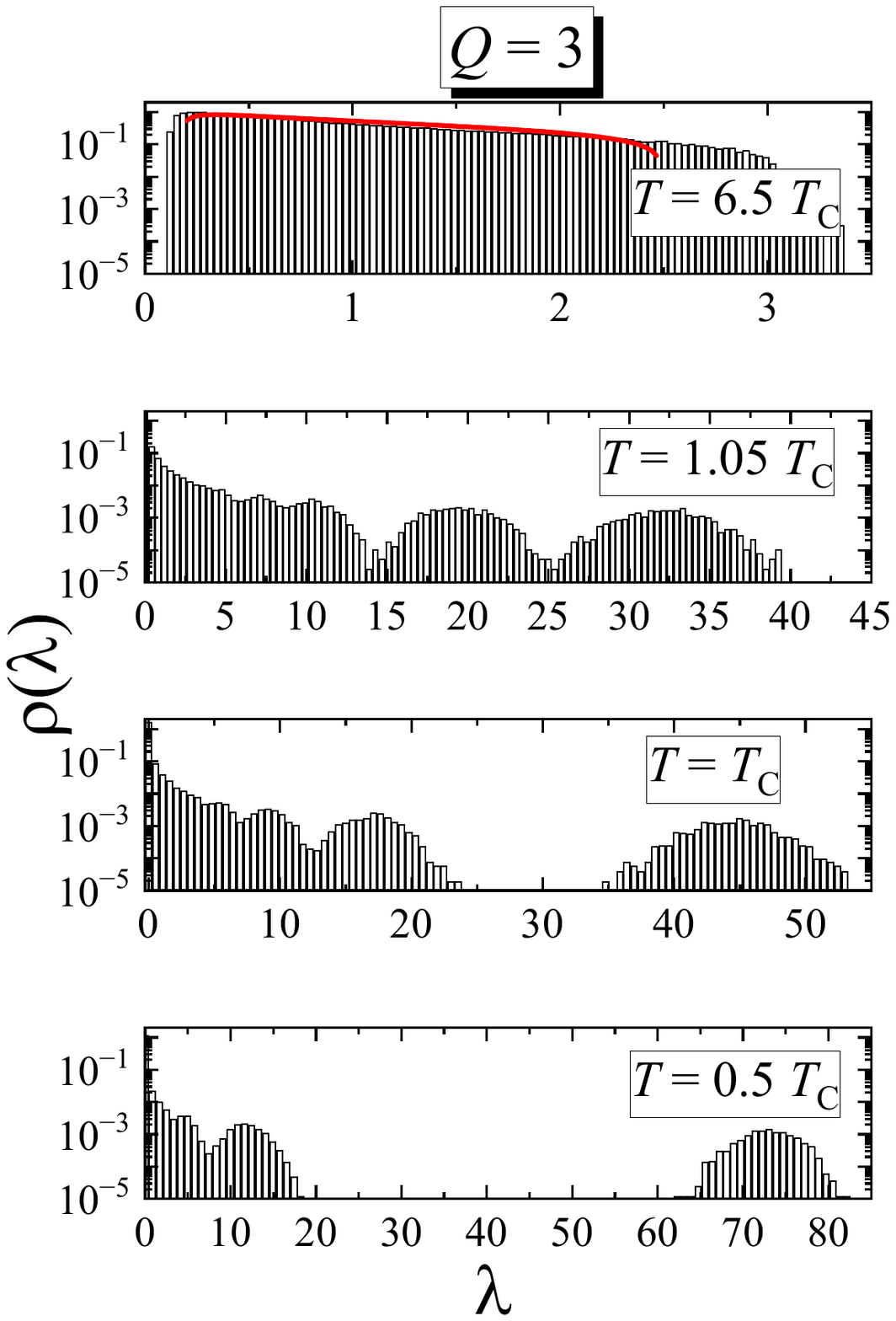}
\end{center}
\caption{Density of eigenvalues for 3-state Potts model. The gap between
eigenvalues disappears after $T>T_{C}$ as previously observed for the Ising
model (see \protect\cite{RMT2023}). A good approximation with
Marchenko-Pastur law (red continuous curve) is observed for high
temperatures. }
\label{Fig:density_potts3}
\end{figure}

\begin{figure}[tbp]
\begin{center}
\includegraphics[width=1.0\columnwidth]{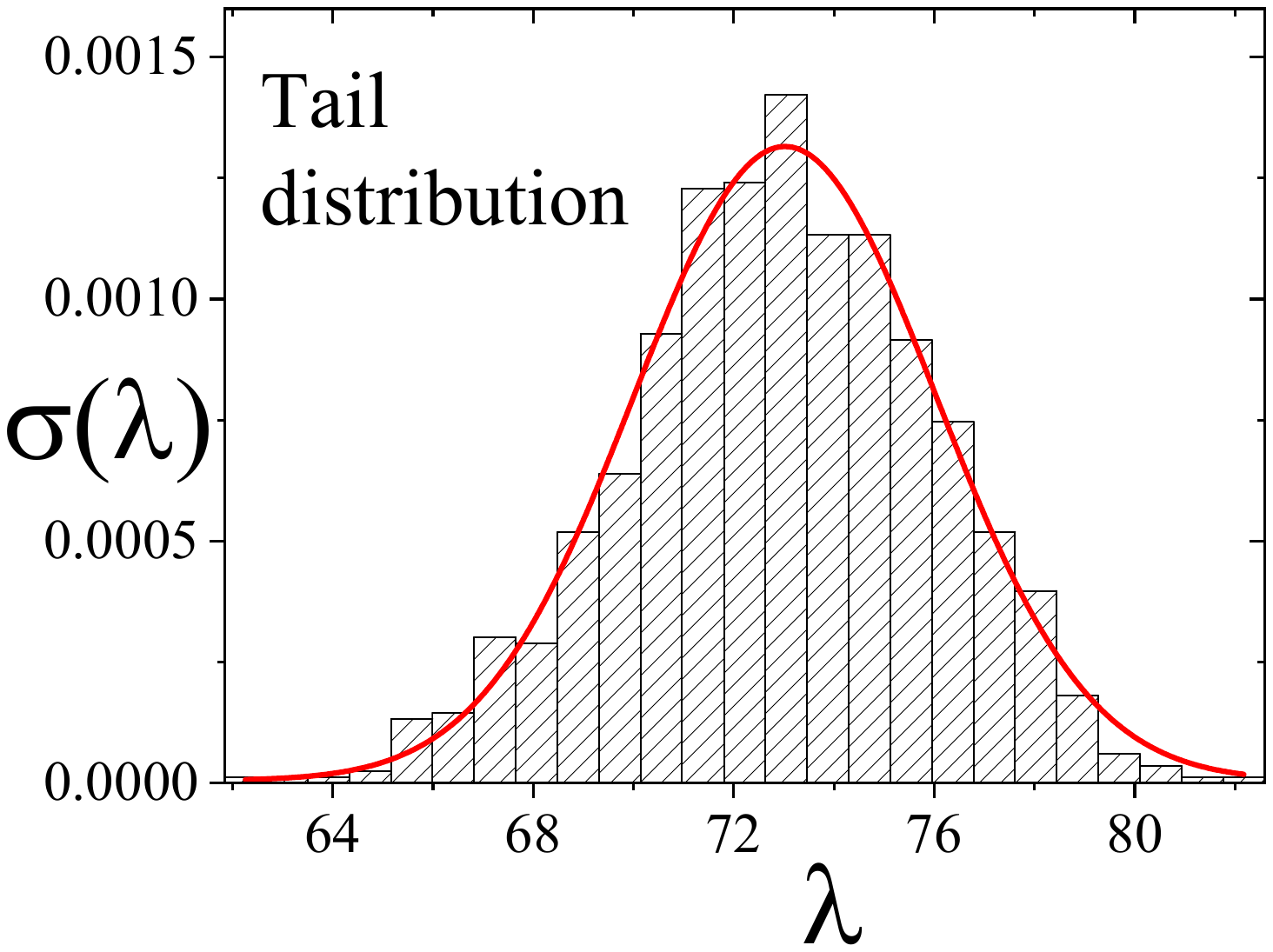}
\end{center}
\caption{The Gaussian behavior stems from the eigenvalues that deviate from
the bulk in Figure \protect\ref{Fig:density_potts3}. \textbf{The eigenvalues
are associated with scenarios characterized by a favorable alignment of the
spins.}}
\label{Fig:gaussian}
\end{figure}

It is important to observe the Gaussian behavior centered at $\lambda =73$\
in Figure \ref{Fig:density_potts3}, which occurs at $T=T_{C}/2$. This is
better highlighted in a linear scale (see Figure \ref{Fig:gaussian}).

The phenomenon is intriguing because there is a single eigenvalue in each
sample that departs from the bulk to create this Gaussian distribution -- an
outlier that indeed signifies a favorable alignment of the spins.

\begin{figure}[tbp]
\begin{center}
\includegraphics[width=1.0\columnwidth]{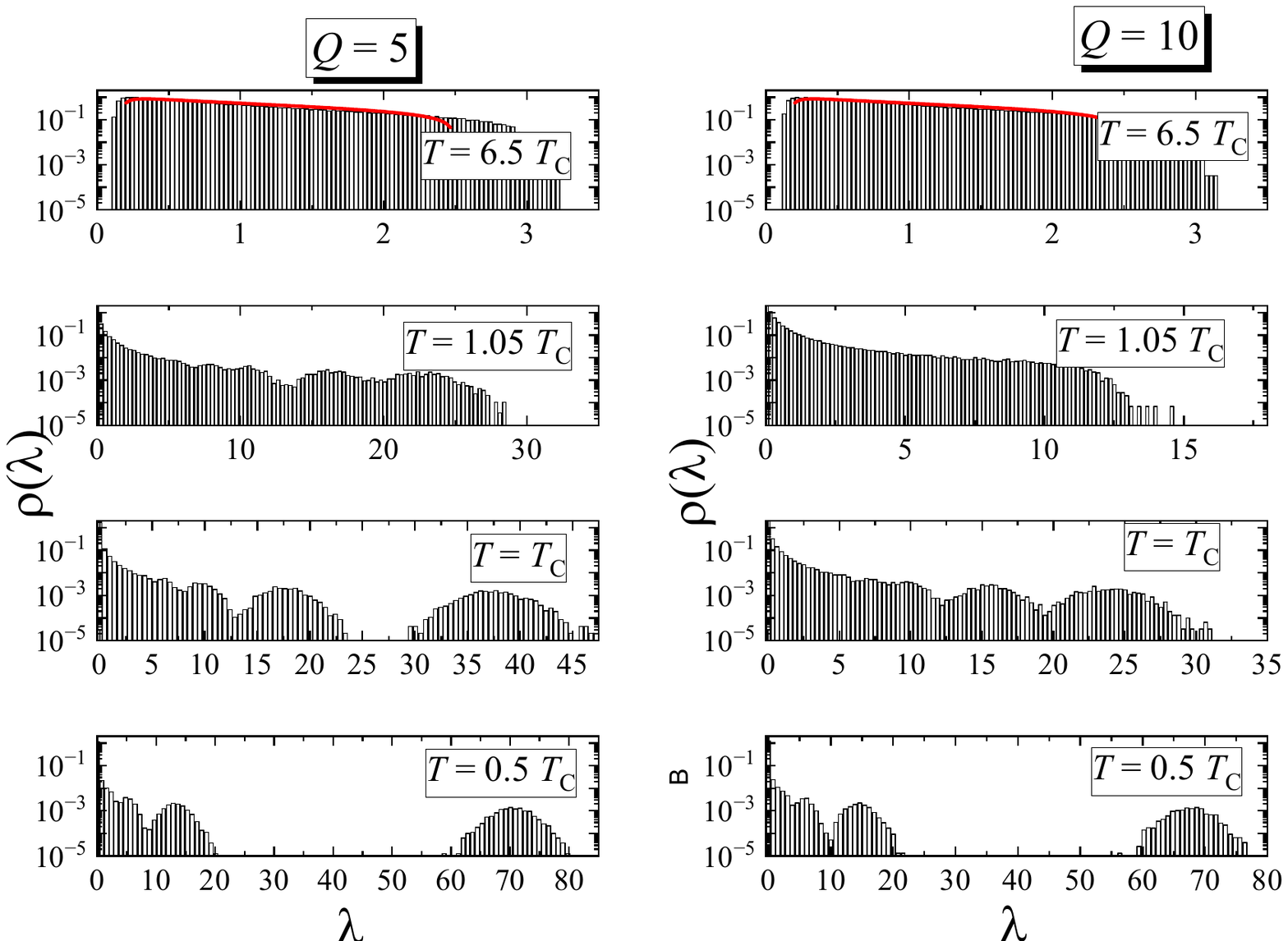}
\end{center}
\caption{Density of eigenvalues for the 5-state and 10-state Potts models.
The gap between eigenvalues disappears even at $T=T_{C}$\ for $Q=10$, which
does not occur for $Q=5$. It is evident that $Q=5$ exhibits characteristics
of a weak first-order point, closely reflecting the critical behavior
observed when $Q<5$.}
\label{Fig:qeq5e10}
\end{figure}

It is interesting to investigate how the density of eigenvalues behaves for $%
Q=5$\ (weak first-order) and $Q=10$\ (strong first-order). Our simulations
in this case lead to results that can be observed in Figure \ref{Fig:qeq5e10}%
.

We can observe that the gap between eigenvalues vanishes even at $T=T_{C}$\
for $Q=10$, which does not occur for $Q=5$. In the case of $Q=5$, it
represents a weak first-order point and mirrors the critical behavior, much
like when $Q=5$.

However, how this phenomenon manifests in the fluctuations of the
eigenvalues is an intriguing question. When conducting simulations for $%
Q\geq 5$, a range associated with first-order transition points, The
observed pattern bears a striking resemblance to the outcomes when $Q\leq 4$%
. Of course, there will be some deviations, which slightly diminish our
initial surprise.

To better understand the mechanism responsible for this characteristic gap
closure, we aim to demonstrate that the phase transitions of the simulated
magnetic system are intricately mirrored by the fluctuations of eigenvalues.
It is evident that these fluctuations are influenced by the presence of
these gaps.

\begin{figure}[tbp]
\begin{center}
\includegraphics[width=0.8%
\columnwidth]{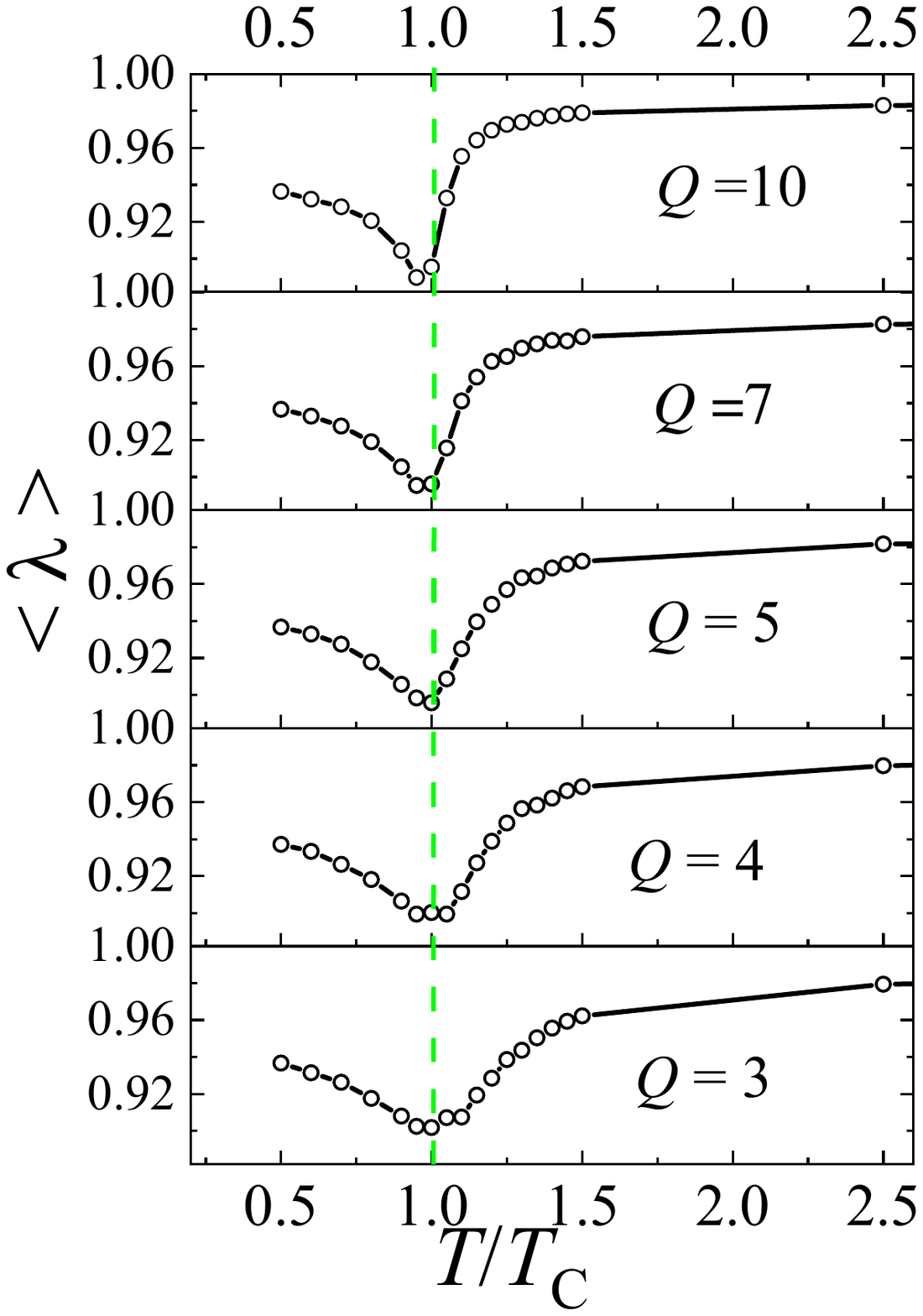}
\end{center}
\caption{Averaged eigenvalue as function of $T/T_{C}$ for $Q=3,4,5,7,$ and $%
10$. }
\label{Fig:average}
\end{figure}

In order to show that, we calculated the first ($k=1$) and second ($k=2$)
moments. The $k$-th moment is given by 
\begin{equation}
\left\langle \lambda ^{k}\right\rangle =\frac{\sum_{i=1}^{N_{bins}}\lambda
_{i}^{k}\sigma (\lambda _{i})}{\sum_{i=1}^{N_{bins}}\sigma (\lambda _{i})}%
\text{.}
\end{equation}%
Fig. \ref{Fig:average} shows the first moment $\left\langle \lambda
\right\rangle $\ as function of $T/T_{C}$. The dispersion defined as $%
\left\langle \left( \Delta \lambda \right) ^{2}\right\rangle =\left\langle
\left( \lambda -\left\langle \lambda \right\rangle \right) ^{2}\right\rangle 
$ as function of $T/T_{C}$ for $Q=3,4,5,7,$ and $10$, is shown in Fig. \ref%
{Fig:dispersion}. 

\begin{figure}[tbp]
\begin{center}
\includegraphics[width=1.0%
\columnwidth]{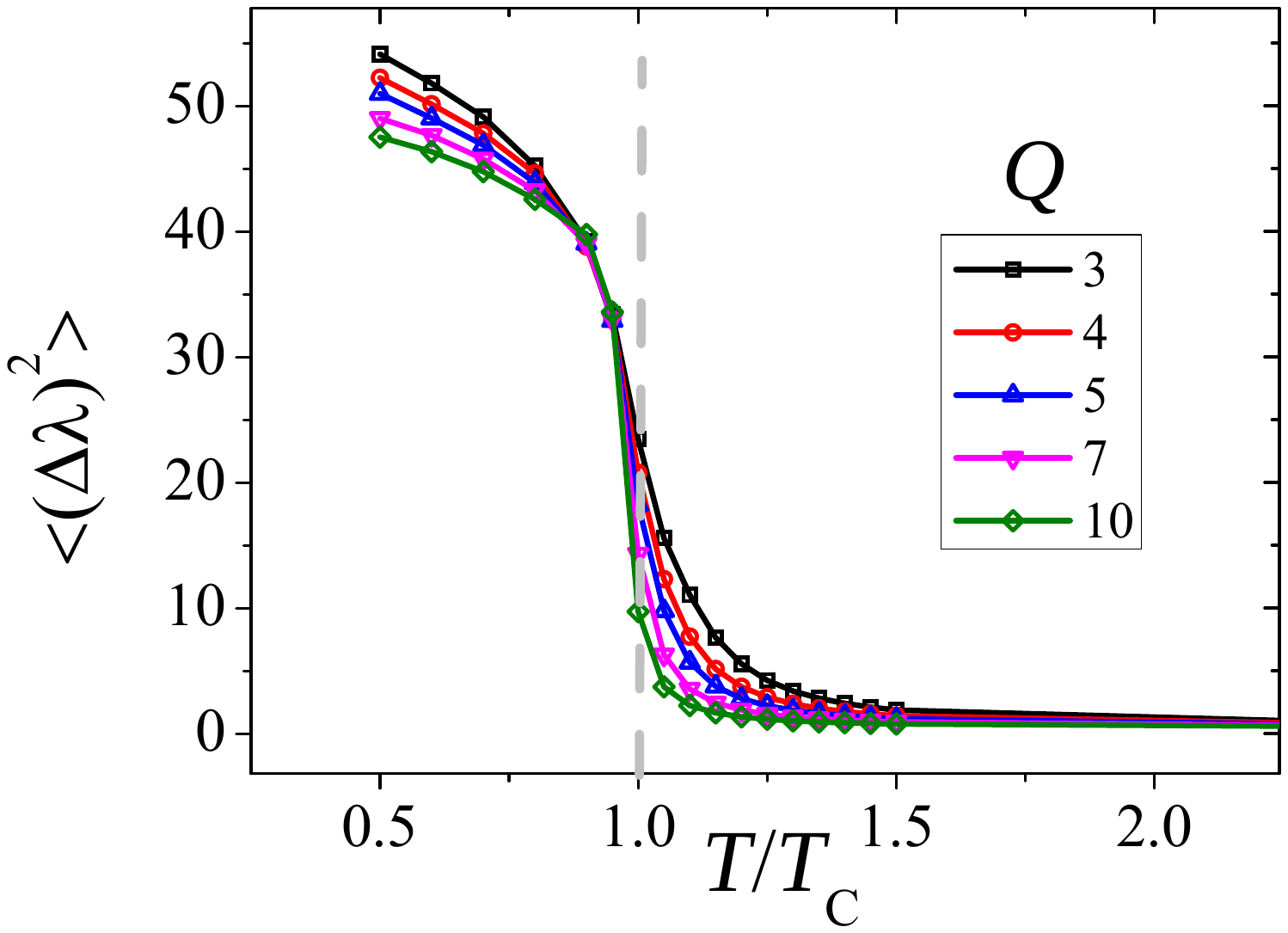}
\end{center}
\caption{Dispersion of the eigenvalue as function of $T/T_{C}$ for $%
Q=3,4,5,7,$ and $10$. }
\label{Fig:dispersion}
\end{figure}

Here the result is really interesting since the average shows a minimum
value in the proximity of critical temperature ($T/T_{C}=1$) for all studied
values of $Q$\ as shown in (Fig. \ref{Fig:average}) except by $Q\geq 7$\
that are stronger first-order points. On the other hand, these first-order
transition points (even for $Q\geq 7$) are in good agreement with the
inflexion points at the same phase transitions that are observed in the
eigenvalue dispersion shown in (Fig \ref{Fig:dispersion}). We have intention
that assert that method can be applied to other strong first order
transition and it deserves a better future exploration.

These results suggest that the eigenvalue spectra respond to the
thermodynamic properties of the correspondent magnetic system. To scrutinize
our results, it is interesting to check the first and second derivatives of
this dispersion. Just for convenience, we have taken the derivative of the
negative of dispersion. The main plot shown in Fig. \ref%
{Fig:Derivative_negative_dispersion} corresponds to the result of the
derivative with the previous interpolation by splines. The inset plot is
without performing the splines.

\begin{figure}[tbp]
\begin{center}
\includegraphics[width=1.0%
\columnwidth]{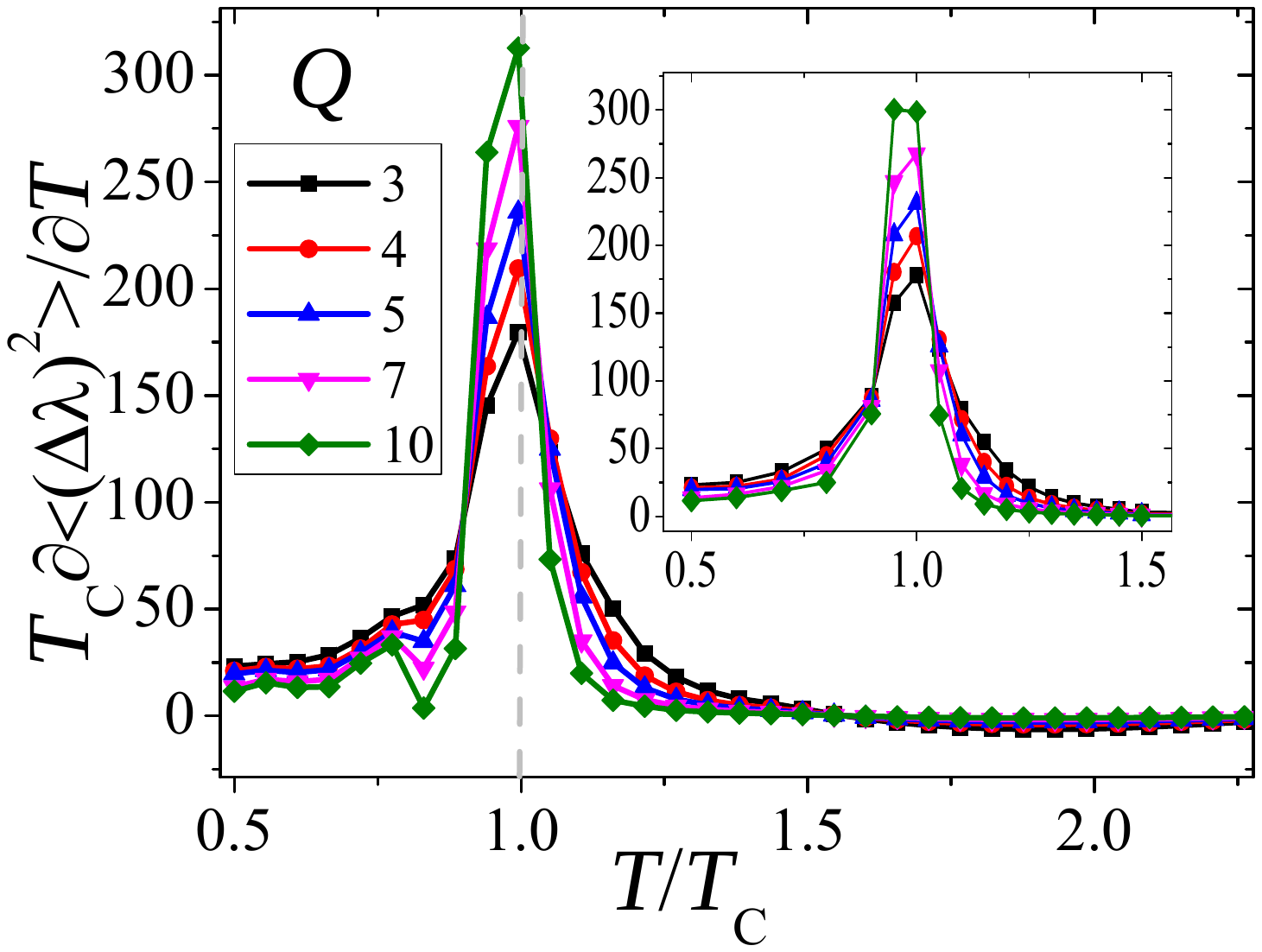}
\end{center}
\caption{Derivative of negative of the dispersion as function of $T/T_{C}$
for $Q=3,4,5,7,$ and $10$. The main plot is the result using splines before
performing the derivative. The inset plot corresponds to the same plot not
using the splines. }
\label{Fig:Derivative_negative_dispersion}
\end{figure}
We observe a maximum at $T=T_{C}$, at most a draw\ for $Q=10$ when we do not
use interpolation by splines. It is important to remember that we have a
strong first-order transition point in that case.

\begin{figure}[tbp]
\begin{center}
\includegraphics[width=0.8%
\columnwidth]{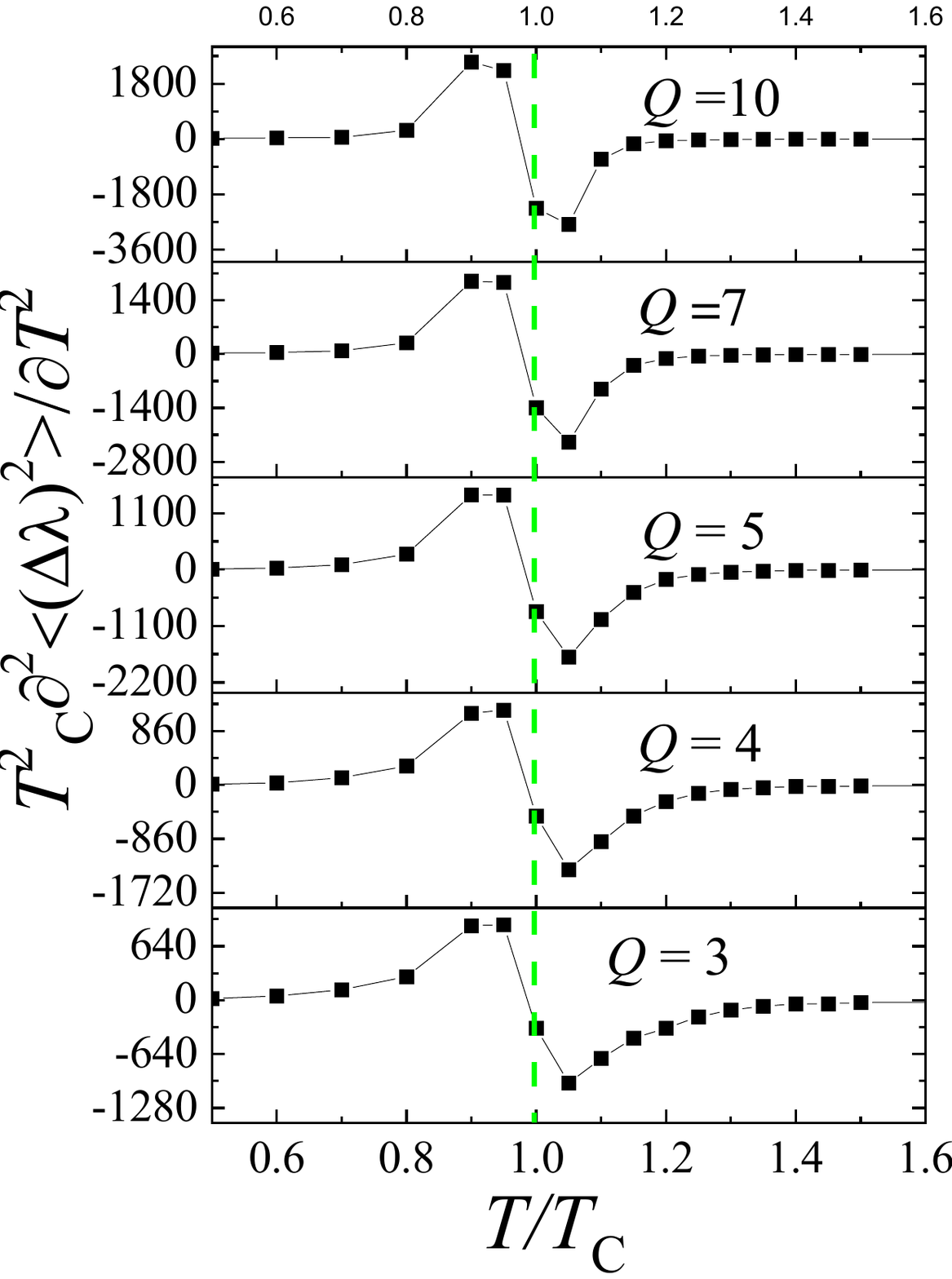}
\end{center}
\caption{Second derivative of dispersion: $T_{C}^{2}\frac{\partial ^{2}}{%
\partial T^{2}}\left\langle \left( \Delta \protect\lambda \right)
^{2}\right\rangle $ as function of $T/T_{C}$ for $Q=3,4,5,7,$ and $10$. As $%
Q $ increases, the discontinuity is numerically more significant, which
makes sense since the order of transition increases.}
\label{Fig:derivative_of_derivative}
\end{figure}

In order to show that the method is definitively a good identifier of
transition points, independent of the transition order, we compute $T_{C}^{2}%
\frac{\partial ^{2}}{\partial T^{2}}\left\langle \left( \Delta \lambda
\right) ^{2}\right\rangle $. We can observe the discontinuity of this
quantity occurring exactly at transition point temperatures, as can be
observed in Fig. \ref{Fig:derivative_of_derivative}.

Up to this point, we gave numerical evidence that the thermodynamics of the $%
Q-$state Potts model is dictated by the fluctuations of the eigenvalues
uniquely imposed by the correlation held in the Wishart matrices. Given the
role played by these correlations in the thermodynamics of the system, we
need to see what happens to the correlation distribution as $Q$ increases
for different critical temperatures.

\subsection{Checking of the histograms of correlation}

\begin{figure*}[tbp]
\begin{center}
\includegraphics[width=1.0\columnwidth]{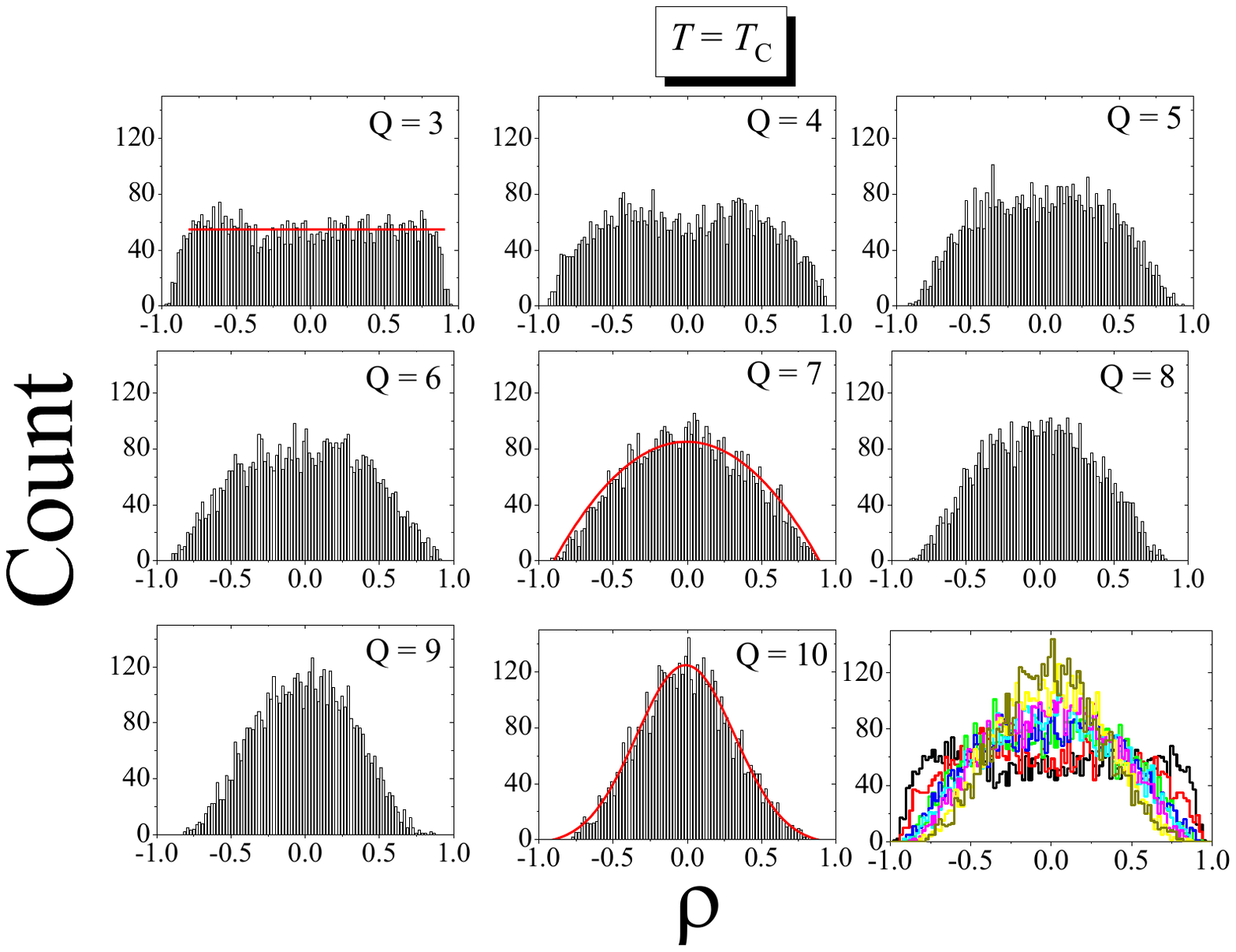}
\end{center}
\caption{Histograms of the correlation for $T=T_{C}$. Here it is interesting
to observe broad histograms distributed in the range $-1 < \protect\rho < 1.$
are observed. For $Q=3$ we almost have a uniform distribution of
correlations. Such distribution is continuously curved until it arrives at a
quadratic function ($Q=7$) which already is a first-order point. For $Q=10$
one has a wide gaussian. The last plot presents all previous plots jointly
for comparison.}
\label{Fig:Correlations_T_EQ_TC}
\end{figure*}

Since we have computed the correlations $\rho _{ij}=\frac{\left\langle
\mathcal{M}_{i}\mathcal{M}_{j}\right\rangle -\left\langle \mathcal{M}_{i}\right\rangle \left\langle
\mathcal{M}_{j}\right\rangle }{s _{i}s _{j}}$
between different pairs of magnetization series, where 
$s _{i}=\sqrt(\left\langle\mathcal{M}_{i}^2\right\rangle-\left\langle\mathcal{M}_{i}\right\rangle^2)$ , for $i>j$, and we perform a histogram of these elements that are identically equal $\mathcal{G}_{ij}$ and here the interchange between symbols $\mathcal{G}_{ij}$ by $\rho_{ij}$ is solely for convenience. We look at situations $T<T_{C}$, $T=T_{C}$ and $T>T_{C}$,
exactly in the same conditions that we used to obtain the eigenvalues of $%
\mathcal{G}$. The results are summarized and reported in Figs. \ref%
{Fig:correlations_T_GT_TC}, \ref{Fig:Correlations_T_EQ_TC}, and \ref%
{Fig:Correlations_T_LT_TC}.

\begin{figure}[tbp]
\begin{center}
\includegraphics[width=1.0\columnwidth]{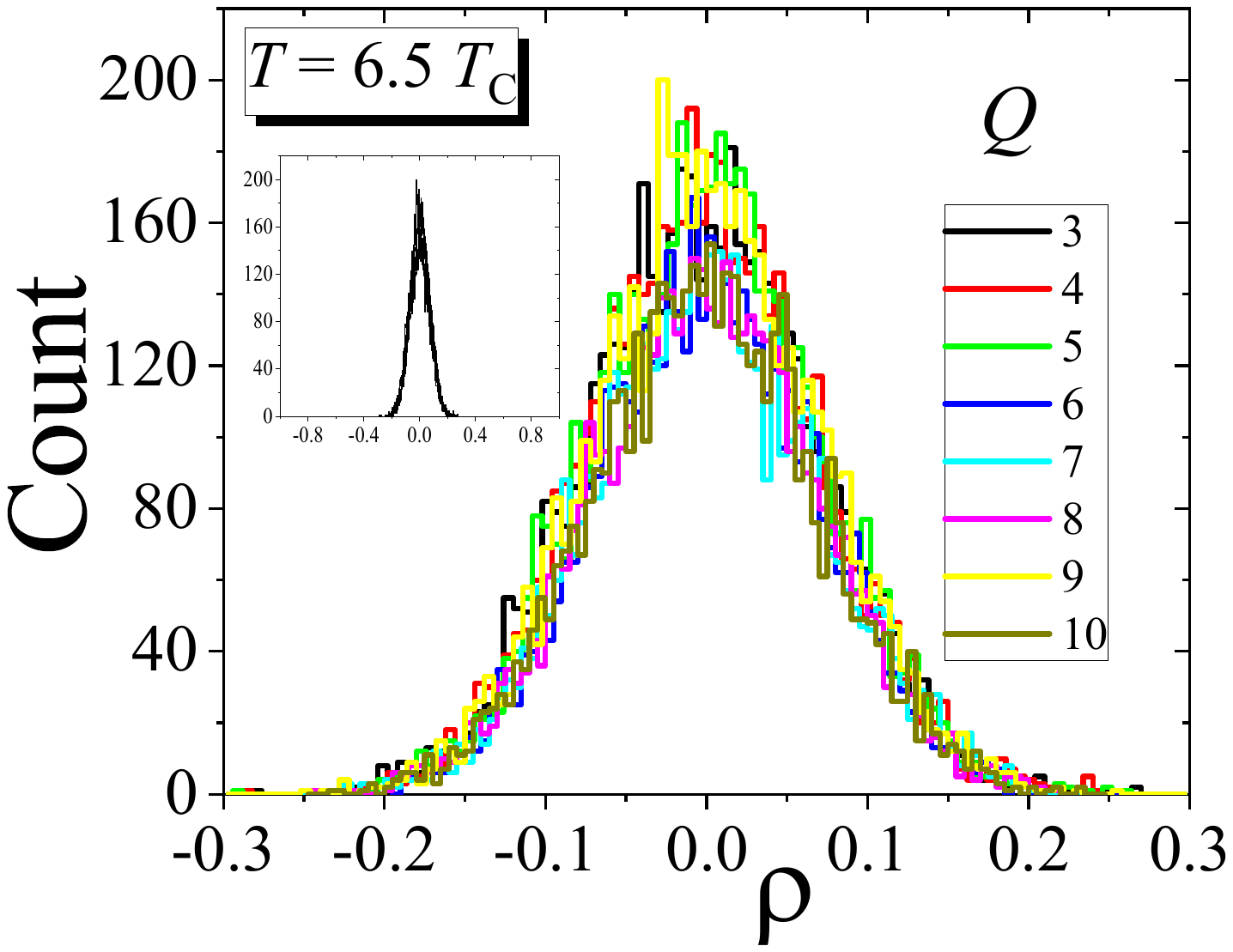}
\end{center}
\caption{Histogram of correlations for $T>T_{C}$. One observes a gaussian
behavior for all $Q-values$. The inset plot presents the same result
considering a large range of correlations by showing histograms narrow
indeed, around $\protect\rho=0$, which in practice means to be uncorrelated}
\label{Fig:correlations_T_GT_TC}
\end{figure}

\begin{figure*}[tbp]
\begin{center}
\includegraphics[width=1.0\columnwidth]{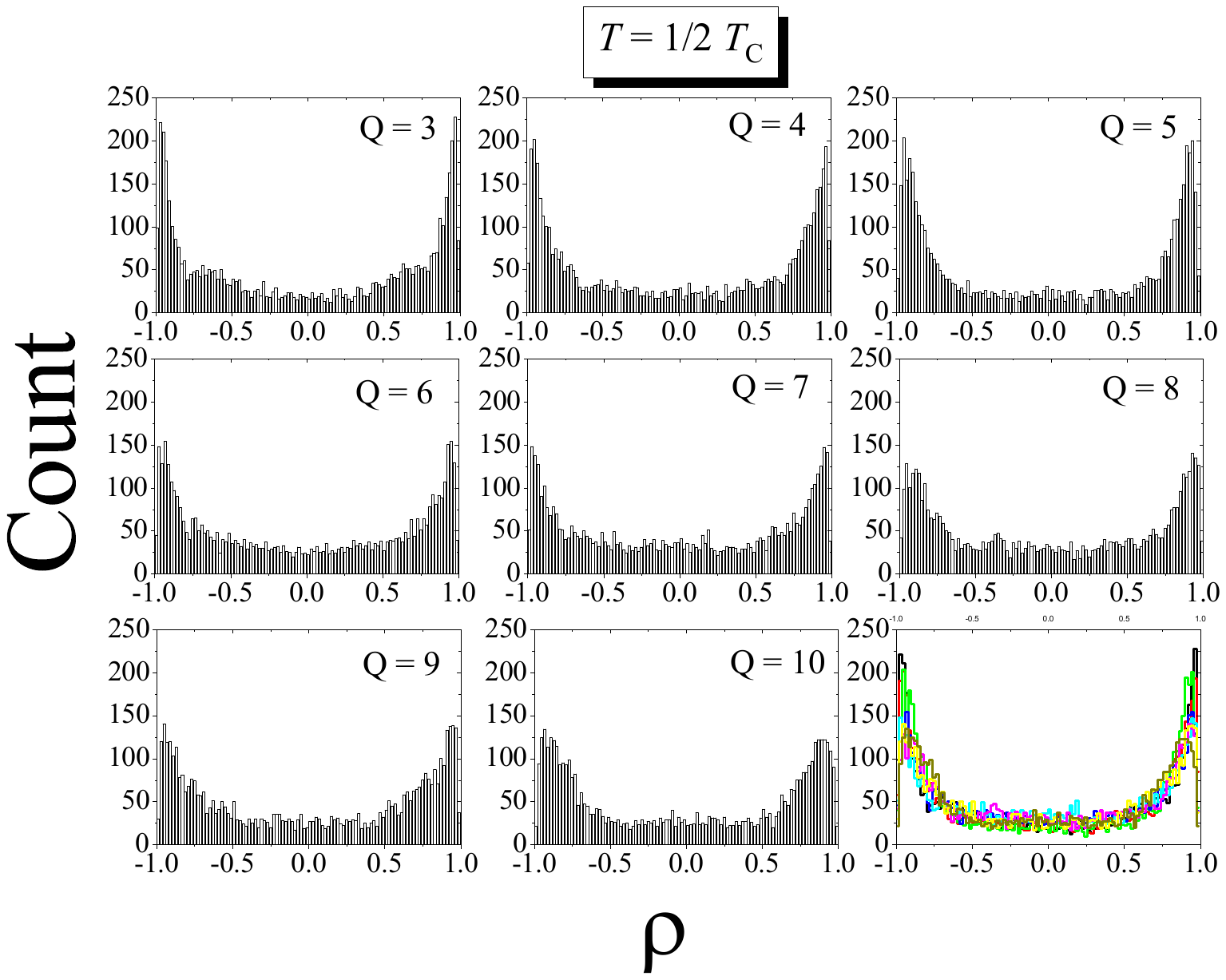}
\end{center}
\caption{Histograms of the correlation for $T<T_{C}$. It is interesting to
observe that independently of $Q$-value, one has similar behavior, double
peak at $\protect\rho\approx1$ and $\protect\rho\approx-1$. The Last plot presents all
previous histograms in the same figure or a comparison}
\label{Fig:Correlations_T_LT_TC}
\end{figure*}

We see that for high temperatures, as shown in Fig. \ref%
{Fig:correlations_T_GT_TC}, the system that starts with a magnetization at $%
T\rightarrow \infty $, keeps this same behavior. We can observe that for all
values of $Q$, the gaussians are narrow exactly as expected. However, for $%
T=T_{C}$, Fig. \ref{Fig:Correlations_T_EQ_TC} the system responds and
acquires orientation so that the correlation follows a broad histogram with $%
\rho$ in the range from $-1$ to $+1$ with null average. The most interesting
point is the changing of this broad histogram that is almost a uniform
distribution for $Q=3$, going throughout a parabolic distribution (see the
already first-order point $Q=7$) and finally becoming a spread gaussian for
the strong first-order point ($Q=10$).

We have to emphasize the possible "indetermination" for $Q=4$, perhaps
indicating the existence of a marginal operator for this Potts model (see 
\cite{Swendsen,Kadanoff,Silva2004}). We can finally see that for $T<T_{C}$,
shown in Fig. \ref{Fig:Correlations_T_LT_TC}, the system really acquires a
strong ordering characterized by a high correlation with two pronounced
peaks in $\rho \approx-1$ and $\rho \approx1$. An observation is that the peaks are
higher for the second-order points. An inevitable conclusion is that the
transition from a two-peaked shape to a broad gaussian-like distribution for
the correlation at $T=T_C$ between eigenvalues probably has a direct
relation to a spectral interpretation of the thermodynamics of systems that
present first and second-order transitions. Our approach to this kind of
problem can bring new ways to deal with the thermostatistics of several
physical systems.

\section{Summaries and Conclusions}

\label{Sec:Conclusions}

We studied the spectral properties of the so-called Wishart matrices built
from the time evolution of spins systems. The spectra fluctuations mimic the
system's thermodynamics independent of the order of the phase transition. We
used the Potts model as our test bed to assess whether the approach
accurately characterizes both the second and weak first-order points.
Additionally, this model allows us to investigate how the method performs in
the presence of a few higher first-order points.

The phase transition occurs because the distribution of correlations changes
from a two-peaked shape at low temperatures to a gaussian-like distribution
for temperatures above the critical one.

From the point of view of the density of eigenvalues of Wishart matrices, we
observe gaps that appear and are \textquotedblleft healed\textquotedblright\
at proximities of transition points (continuous or discontinuous ones).

Furthermore, we conducted a preliminary investigation using random matrices
artificially generated with a fixed correlation $\rho $ between pairs of
columns in the matrix. The findings suggest that the density of eigenvalues
undergoes a transition: from a scenario with a pronounced gap between the
eigenvalues ($\rho \approx >0.9$), which diminishes as $\rho $\ decreases,
until it reaches a state characterized by a unified bulk of eigenvalues at $%
\rho \approx 0$. These nuances, albeit somewhat pedagogically, provide
insight into what the fluctuations of the eigenvalues might reveal in the
Potts model. This is due to a direct relationship between temperature and
correlation in magnetic systems.

Our results also demonstrate that the fluctuation spectra of these matrices
are analogous to the direct response of the thermodynamic system, providing
insights into the system's criticality and weak first-order points. Further
investigations into first-order points, including other models such as the
Blume-Capel or Metamagnetic model, will be a focus of our future work, since
the deviations occur for the higher first order points when we analyzed $%
\left\langle \lambda \right\rangle \times T/T_{C}$. Additionally, the
application of this method to long-range spin systems (not only that ones in
mean-field \cite{RMT-2}) holds promise and warrants exploration in future
studies.

\begin{figure*}[tbp]
\begin{center}
\includegraphics[width=1.0\columnwidth]{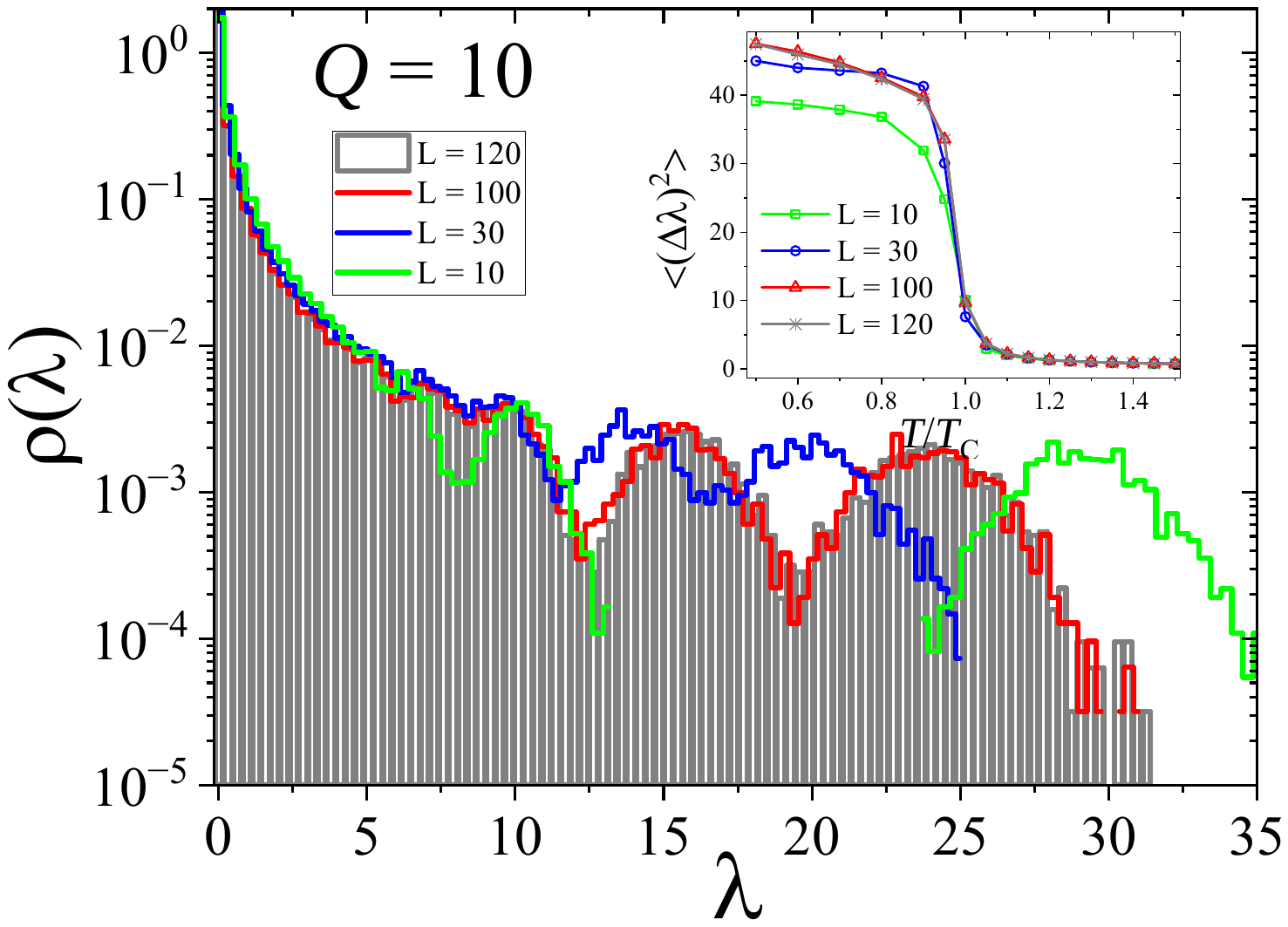}
\end{center}
\caption{Finite size scaling effects on the density of eigenvalues for $Q=10$. 
Notably, there is negligible variation between $L=100$ and $L=120$. The inset 
plot illustrates the variance across these distinct lattice sizes. Likewise, comparable observations are noted for the two largest lattices.} 
\label{Fig:FSS}
\end{figure*}

It is crucial to emphasize that the conclusions drawn in this paper are highly reliable and not influenced by finite size scaling (FSS) issues or similar factors. For instance, readers should note that in Figure \ref{Fig:qeq5e10}, the gap vanishes for $Q=10$ but not for $Q=5$, potentially due to FSS effects. However, as depicted in Figure \ref{Fig:FSS}, this assumption does not hold true.

In this figure, it's evident that there is negligible difference in the density of states between $L=100$ and $L=120$. To dispel any doubts regarding the adequacy of our chosen lattice size for the method, the inset plot within the same figure depicts the eigenvalue variance across these identical lattice sizes. Similarly, no discernible visual distinctions are observed between these two lattice sizes. 
   
Last, but not least, we emphasize that the method presented in this paper
works with lower computational complexity compared to other similar methods.
It diagonalizes matrices of size $O(N_{sample})$, where $N_{sample}$\ is
100, for example, in this contribution. This is in contrast to other methods
that require diagonalizing matrices of size $O(L^{d})$, where $L$\ is the
linear dimension of the system and $d$\ is its dimensionality. For a system
with $L=100$\ and $d=2$, this would mean diagonalizing matrices of size $%
10^{4}\times 10^{4}$.

\bigskip

\end{document}